\documentclass[]{tPHM2e}

\usepackage{subfigure}
\usepackage{amsmath}
\usepackage{graphics}
\usepackage{rotating}

\begin{document}
\doi{10.1080/14786435.20xx.xxxxxx}
\issn{1478-6443}
\issnp{1478-6435}
\jvol{00} \jnum{00} \jyear{2011} 

\title{Visualization of nano-plasmons in graphene}

\author{Hari Dahal,$^{\rm a,b}$
Rodrigo A. Muniz,$^{\rm c}$ 
Stephan Haas,$^{\rm c}$
Matthias J. Graf,$^{\rm a}$\thanks{$^\ast$Corresponding author. Email: graf@lanl.gov}
and
Alexander V. Balatsky$^{\rm a,b}$
\\\vspace{6pt}  $^{\rm a}${\em{Theoretical Division, Los Alamos National Laboratory, Los Alamos, NM 87545, USA}}
\\
$^{\rm b}${\em{Center for Integrated Nanotechnologies, Los Alamos National Laboratory, Los Alamos, NM 87545, USA}}
\\
$^{\rm c}${\em{Department of Physics and Astronomy, University of Southern California, Los Angeles, CA 90089}}
}

\date{\today}

\maketitle

\begin{abstract}

We study localized plasmons at the nanoscale (nano-plasmons) in graphene.
The collective excitations of induced charge density modulations in graphene are drastically changed in the vicinity of a single impurity compared to graphene\rq{}s bulk behavior.
The dispersion of nano-plasmons depends on the number of electrons and the sign, strength and size of the impurity potential. 
Due to this rich parameter space the calculated dispersions are intrinsically multidimensional
requiring an advanced visualization tool for their efficient analysis, which can be achieved with parallel rendering.
To overcome the problem of analyzing thousands of very complex spatial patterns of nano-plasmonic modes, 
we take a combined visual and quantitative approach to investigate the excitations on the two-dimensional graphene lattice.
Our visual and quantitative analysis shows that impurities trigger the formation of localized plasmonic excitations of various symmetries. 
We visually identify dipolar, quadrupolar and radial modes, and quantify the spatial distributions of induced charges.
\bigskip

\begin{keywords}graphene; plasmon; tight-binding model; impurity; quantitative visualization
\end{keywords}\bigskip

\end{abstract}

\section{Introduction}

Graphene is a single-layer atomically thin allotrope form of carbon, where  carbon atoms are arranged in a honeycomb lattice.
Owing to its single-atom thickness and unique electronic properties
it is a promising material for technological use \cite{geim2007}.
Although graphene was originally thought to be a pure system,
recent scanning tunneling microscopy work shows that it is a disordered system \cite{brar2007}.
It has lattice defects, vacancies, ripples, dislocations, magnetic impurities, etc. 
Particularly there are charge puddles caused by chemical adsorption or molecules trapped between the graphene lattice and substrate \cite{zhang2009}.
Previous studies of impurities have focused on understanding their effects on the ground state electronic properties of graphene \cite{stauber2005, katsnelson2008, kumazaki2007, pereira2007, bena2008, wehling2007, wehling2009, brar2007, mallet2007, tan2007, ni2009, deshpande2009}.
However, an equally important aspect is the study of the consequences of these charged impurities on the dielectric response for future electronic devices.

Several studies explored plasmonic excitations in pristine graphene \cite{wang2007,hwang2007,gangadharaiah2009,hill2009}.
So far almost no work focused on localized plasmons (nano-plasmons) in graphene.
These excitations have the ability to concentrate an electric field (visible photons) at the nanoscale regime, 
thereby beating the diffraction limit.
The potential technological applications of this effect are vast.
For example, it is possible to improve solar cell efficiency by increasing the cross section of the absorbing material without requiring a thick layer of it \cite{atwater2010}.
Another potential application is the new sub-field of photonics called plasmonics, where plasmons substitute photons in the information transport extending photonics into the nanoscale world \cite{ozbay2006}.
In biology the potential applications include labeling and imaging, optical sensing and photo-therapy treatment of cancer \cite{boisselier2009}.

Here we study nano-plasmons, that is, localized plasmonic excitations in pristine and disordered graphene.
When plasmons are localized the translational invariance of the lattice is lost, hence analytical calculations become impractical.
A self-consistent quantum-mechanical approach to nano-plasmons in graphene in the presence of impurities has been developed, which accounts for the non-locality of the dielectric response function \cite{ilya2006,rodrigo2009}.
Here we use a generalized version of this approach, where the polarization operator is diagonalized, providing information about individual plasmonic excitations, including spatial profiles and local spectral density of states.
The calculation is performed for a single impurity at the center of the carbon hexagon.
We study the impurity effect as a function of the sign, strength, and size of the impurity potential, and the doping level of graphene.
The interactive use of three-dimensional (3D) visualization allows us to identify readily which plasmonic excitations are localized vs.\ delocalized.
Our visualization of the poles of self-consistently calculated polarization operators shows the existence of nano-plasmons around naturally occurring impurities in graphene, which confirms that graphene is an intrinsic plasmonic material \cite{rodrigo2010}.

The paper is organized as follows: In Section 2, we introduce the lattice model used to calculate the plasmonic excitations in graphene.
In Section 3, we describe the needs for an advanced 3D visualization tool to analyze complex multidimensional data sets and
discuss the computational requirements. The results of the plasmonic modes are presented in Section 4.
In Section 5, we perform a quantitative analysis of the symmetries of the various nano-plasmonic modes and conclude in Section 6.


\section{Lattice model}

The electronic structure of graphene is described by the Hamiltonian $ H = H^0 + V $, where $H^0$ is the kinetic and potential energy, while $V$ is the Coulomb interaction between electrons. The interaction is treated as a perturbation and will be considered later in the Random Phase Approximation (RPA).
The single-particle part of the Hamiltonian on a lattice is
\begin{equation}
H^0 =-t\sum_{<ij>}\left( c_{i}^{\dagger}c_{j} + c_{j}^{\dagger} c_{i}\right) +
\sum_{i}U_{i}c_{i}^{\dagger}c_{i}  +  \mu \sum_{i} c_{i}^{\dagger}c_{i} ,
\label{eqn:hamiltonian}
\end{equation}
where $<ij>$ denotes nearest neighbor sites; $c_{i}^{\dagger}$ and $c_i$ are operators that create and annihilate electrons at site $i$;
$t=2.7$ eV is the tight-binding hopping parameter, and $\mu$ is the chemical potential ($\mu=0$ for undoped graphene).
$U_i =U _{0}\exp \left( \frac{-\left\vert {\mathbf{x}}_i-{\mathbf{x}_0} \right\vert ^{2}}{2\sigma ^{2}} \right)$
is the change in the on-site potential for atom at site $\mathbf{x}_i$ due to the presence of an impurity at $\mathbf{x}_0$. $U_{0}$ is the strength of the
impurity potential and  $\sigma $ determines its spatial spread.
The matrix structure of the non-interacting Hamiltonian $H^0$ is
\begin{equation}
H^0 =
\begin{bmatrix}
\ddots & U_{i-1}+\mu & -t & 0 & \ddots \\
   & -t & U_i+\mu & -t &  \\
\ddots & 0 & -t & U_{i+1}+\mu & \ddots
\end{bmatrix}.
\end{equation}
The  diagonalization of the non-interacting single-particle Hamiltonian $H^{0}$ gives eigenstates $\left\vert \Psi_{\alpha}^{0} \right\rangle$ and eigenvalues $E_{\alpha }^{0}$ which satisfy
$\sum_{j} H^0_{ij} \left\vert \Psi_{\alpha}^{0}\right\rangle_j = E_{\alpha }^{0}\left\vert \Psi _{\alpha}^{0}\right\rangle_i$. Each state has an occupation number given by the Fermi function $n^0_\alpha = [{\exp(-\frac{E^0_\alpha-\mu}{k_B T}) +1}]^{-1}$.

In order to compute the polarization operator due to an induced charge, we define the tensorial matrices $\Delta n$ and $\Delta E$ in the basis of the eigenstates
\begin{eqnarray}
\Delta n_{\alpha \beta ,\gamma \delta } &=&\delta _{\alpha \gamma }\delta
_{\beta \delta }\left( n_{\alpha }^{0}-n_{\beta }^{0}\right)   \notag \\
\Delta E_{\alpha \beta ,\gamma \delta } &=&\delta _{\alpha \gamma }\delta
_{\beta \delta }\left( E_{\alpha }^{0}-E_{\beta }^{0}\right) \text{.}
\end{eqnarray}
Using the superindices $I=\alpha \otimes \beta$ and $J=\gamma \otimes \delta$, which reshape a matrix into a vector in a column-wise fashion so while $\alpha, \beta \in {1,\ldots,n} $ one has $ I \in {1,\ldots,n^2}$, the above tensors can be written as diagonal matrices
\begin{equation}
\Delta n =
\begin{bmatrix}
 n^0_\alpha - n^0_{\beta-1} & 0 & \ddots \\
   0 &  n^0_\alpha - n^0_{\beta} & 0 \\
\ddots & 0 & n^0_\alpha - n^0_{\beta+1}
\end{bmatrix}
\end{equation}
and
\begin{equation}
\Delta E =
\begin{bmatrix}
 E^0_\alpha - E^0_{\beta-1} & 0 & \ddots \\
   0 &  E^0_\alpha - E^0_{\beta} & 0 \\
\ddots & 0 & n^0_\alpha - n^E_{\beta+1}
\end{bmatrix} .
\end{equation}
The polarization operator is a density-density response function. 
Summing the product of non-interacting Green's functions
over Matsubara frequencies $\omega_m$ one obtains the non-interacting polarization operator
\begin{equation}
\begin{array}{rl}
\Pi_{\alpha\beta,\gamma\delta}^{0}\left(\omega\right) & =\sum_{\omega_{m}}G_{\alpha\gamma}^{0}(\omega+i\omega_{m}) G_{\delta\beta}^{0}(i\omega_{m})=\sum_{\omega_{m}} \frac{\delta_{\alpha\gamma}}{\left(\omega+i\omega_{m}-E_{\alpha}^{0}\right)} \frac{\delta_{\delta\beta}}{\left(i\omega_{m}-E_{b}^{0}\right)}\\
 & =\delta_{\alpha\gamma}\delta_{\beta\delta} \frac{n_{\alpha}^{0}-n_{\beta}^{0}}{E_{\alpha}^{0}-E_{\beta}^{0}-\omega}{.}
\end{array}
\end{equation}
Using the superindices $I$ and $J$, one can write $\Pi^{0}$ as the product of diagonal matrices
\begin{equation}
\Pi^{0}\left( \omega \right) = \Delta n \left( \Delta E-\omega I\right) ^{-1} ,
\label{eqn:polariz}
\end{equation}
which again is a diagonal matrix.

Next we include the Coulomb interaction $V$ for electrons on the lattice.
The interaction Hamiltonian is written in terms of creation and annihilation operators in the lattice basis as
\begin{equation}
V = \sum_{ijlm} V_{ij,lm} c_{i}^{\dagger} c_{j} c_{l}^{\dagger} c_{m},
\end{equation}
where the only non-zero components are
\begin{equation}
V_{ij,ij} = V_{ii,jj} = \frac{V_0}{|{\bf r}_i - {\bf r}_j|}.
\end{equation}
We can write $V$ in the basis of the eigenstates, where it will be a full tensorial matrix.
Therefore $V$ can also be represented as a matrix in the superindex space $V_{\alpha\beta,\gamma\delta} \to V_{IJ}$.

Within the RPA calculation the polarization operator of the interacting system is given by
\begin{equation}
\Pi (\omega ) = \Pi^{0}\left( \omega \right) \left( 1-V \Pi^{0}\left(
\omega \right) \right)^{-1} .
\end{equation}
Using Eq.~(\ref{eqn:polariz}) we can write
\begin{equation}
\Pi \left( \omega \right) = \Delta n\left( \Delta E  - V\Delta n - \omega I \right)^{-1} .
\end{equation}
The nano-plasmonic excitations correspond to the poles of the polarization operator.
Hence the modes correspond to charge density oscillations with amplitude 
\begin{equation}
\delta \rho_{I} = \sum_J \Delta n_{IJ} {\phi}_J ,
\end{equation}
such that
\begin{equation}
\left( \Delta E-\omega I-V\Delta n\right) {\phi }=0 .
\label{eqn:condition}
\end{equation}
The matrix $\Delta E-\Delta n V$ is diagonalized giving the eigenvalues $\omega^\lambda$ and eigenvectors $\phi^\lambda_I$ satisfying $\sum_J (\Delta E-\Delta n V)_{IJ} \phi^\lambda_J = \omega^\lambda \phi^\lambda_I$.
The polarization has  poles at $\omega = \omega^\lambda$ for each eigenvalue $\omega^\lambda$ of $\Delta E -\Delta n V$.
One can then find the spatial profile of the mode with energy $\omega^\lambda$ through $\delta \rho^\lambda_{I} = \sum_J \Delta n_{IJ} {\phi}^\lambda_J$.
By going back to the lattice basis and neglecting the orbital overlap we get the local charge oscillation
\begin{equation}
\delta \rho^\lambda_{I} = \delta \rho^\lambda_{\alpha\beta} \to \delta \rho^\lambda_{ij} \simeq \delta \rho^\lambda_{ii}.
\end{equation}
The amplitude of the induced charge density oscillation at site $i$ is the vector $\delta \rho^\lambda_{i} \equiv \delta \rho^\lambda_{ii}$.
Of course the induced charge will oscillate with frequency $\omega^\lambda$, that is, 
$\delta \rho^\lambda_{i}(t) = \delta \rho^\lambda_{i} \exp(-i \omega^\lambda t)$.

We use a standard numerically exact diagonalization method to solve Eq.~(\ref{eqn:condition}). This allows us to find all the poles of the polarization operator and, thus, we 
obtain the complete information of the plasmonic excitations on the graphene lattice. For a lattice having $N$ sites, the matrix size of $\Pi$ becomes $N^4 = N^2 \times N^2$.
The number of eigenmodes scales as $N^2/4 = N/2 \times N/2$ for $N/2$ electron-hole pairs at half-filling.
Therefore the number of data points where the plasmon dispersion needs to be visualized scales as $N^3/4$, which becomes rapidly a big number even for small lattice sizes.


\section{The need for advanced 3D visualization}

Visualization plays an important role in scientific data analysis with a growing need
for being able to visually analyze
complicated data sets in three dimensions (in real space) and more (in parameter space).
The analysis of such data is always complemented by the ability to effectively interact
with the data. Interacting with multidimensional data sets in a 3D environment (stereo display) 
adds an additional advantage to navigating and seeing {\it through} complex data sets. 
Although this advanced visualization feature is not always necessary for the data analysis.
The need for visual analysis of large data sets (larger than the size that a typical computer with a graphics card can process)
is growing. The problem is commonly tackled by using distributed parallel processing on multiprocessor units.
Here we discuss a specific example of analyzing a multidimensional, medium-sized data set (on the order of Megabytes).
The data set has 200,000 to 300,000 points in three dimensional space.
In this paper, we show the efficiency of a small-scale parallel rendering computer cluster as a possible solution.
The data is obtained from a calculation of the spatial distribution of plasmonic excitations (induced charges on lattice sites) in pristine and impure graphene. 
Since visual analysis of data combined with
quantitative analysis provides a better understanding of the data studied, 
we present specific examples for the quantification of visual information using the open-source visualization tool ParaView \cite{paraview}.

We analyze the dispersion of the plasmonic excitation as a function of energy $\omega^\lambda \to \omega$.
Since the graphene lattice is two dimensional, the analysis of the induced charge distribution on individual
lattice sites as a function of energy  $\omega$ adds one more dimension to the visualization problem.
For a given impurity potential, a single data set is already four dimensional (4D) in the hyper-space of variables
(X, Y, $\omega$, $\delta \rho$(X,Y,$\omega$)). The X-Y dimension is fixed by the number of
lattice sites chosen for the calculation. We fixed it to be $N=96$ lattice sites.
The Z-dimension is determined by the number of possible energies $\omega$ of the plasmonic excitations.
For 96 lattice sites the number of unique frequencies $\omega$ is about $2300$, and the number of induced charges 
as a function of frequency to be visualized is about 
$N^3/4\approx 220,000$.
To specify the induced charge oscillation at each lattice site, we need one scalar value and three coordinates.
Further, we need at least two colors to represent the polarity of the induced charge. In three dimensions it is necessary
to use a 3D glyph to represent the scalar value (induced charge). We use a "sphere glyph" having at least 10 
points of resolution for angle $\theta$ and 10 points for angle $\phi$ \cite{paraview_book, paraview_guide}. In the XML unstructured grid data file format 
\cite{vtk_userguide, paraview_book, paraview_guide, paraview_help},
for which the visualization tool ParaView has a reader to input data, the size of the raw data file is about 10 Megabytes. 
After using the 3D sphere glyphs, to represent the induced charge distribution,
the size of the data set to be processed becomes about 1.3 Gigabytes $\sim 2 \times 10^2 \times 10$ Megabytes.
So the memory required by the computer to process the data becomes more than 100 times larger than the size of the original raw data set. 
One can use a single-processor computer with a high-end graphics card to analyze and post-process a couple of such data sets, but many data sets of this size cannot be efficiently processed, let alone attempting a real-time or interactive data analysis. 
Hence using a single-processor computer is neither time efficient nor practical and parallel rendering on multiple cores or CPUs (central processing units) or multiple GPUs (graphical processing units) becomes mandatory. In the remainder of this study we demonstrate how distributed parallel rendering on multiple CPUs can solve the problem of large data sets.

\begin{figure}
  \centering
 \begin{tabular}{|c|c|}
 	\hline
  \includegraphics[width=5cm, height=5cm]{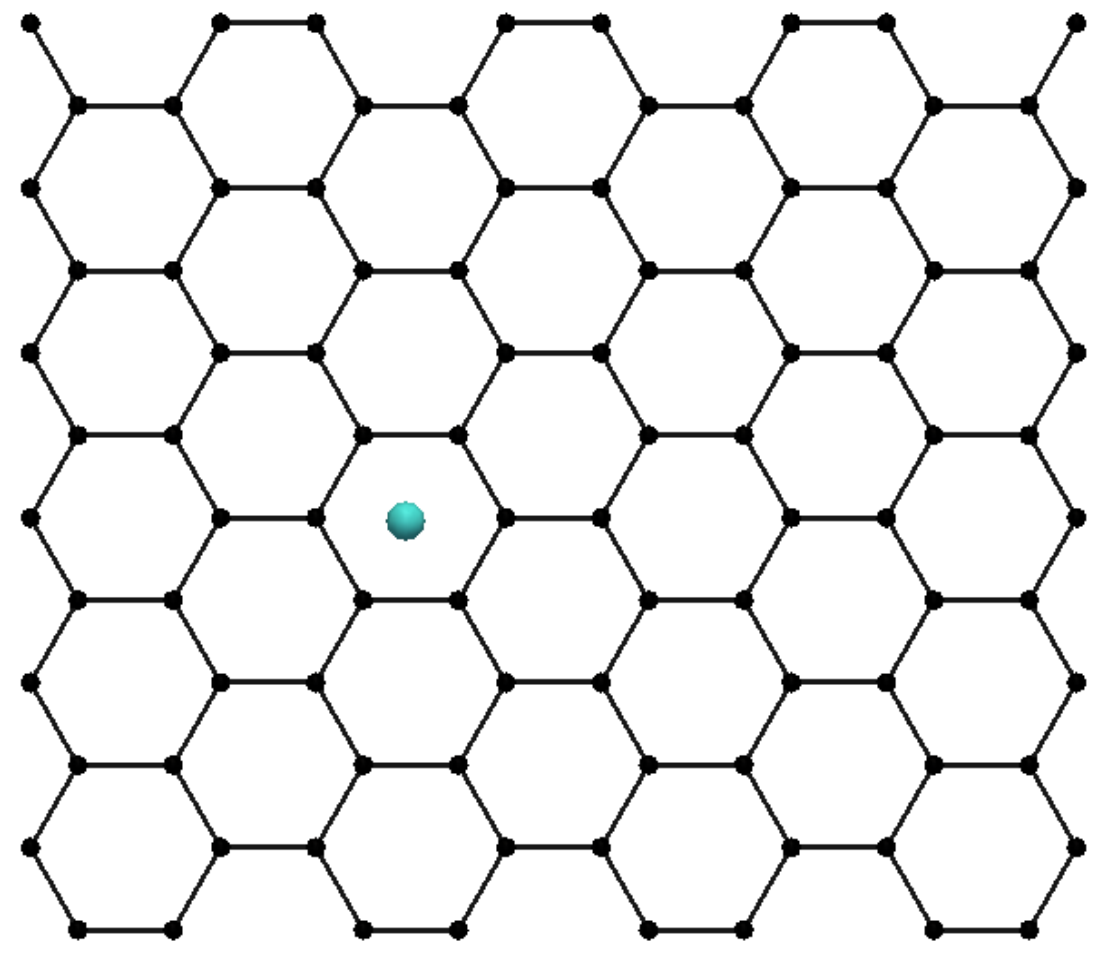} &
  \includegraphics[width=5cm, height=5cm]{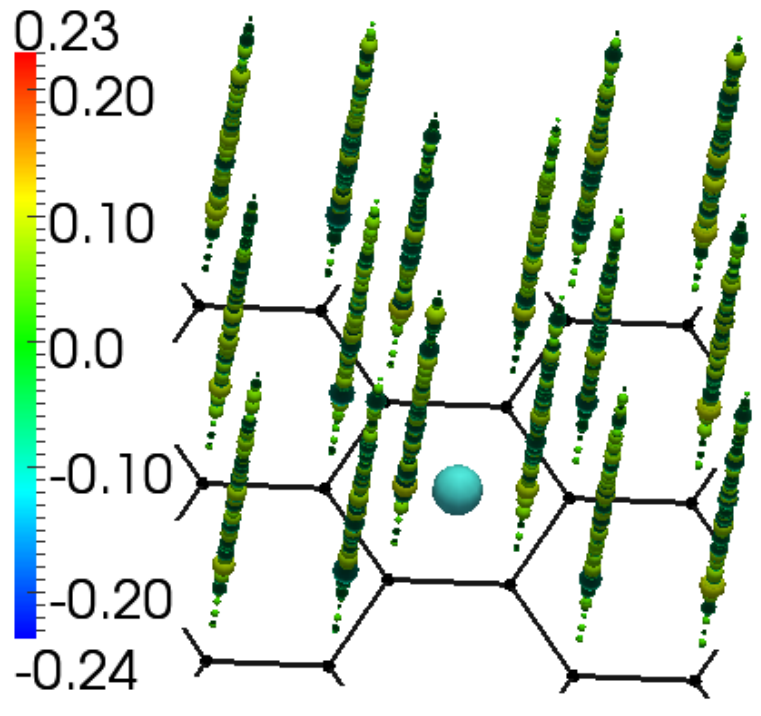}\\
   	\hline
    \end{tabular}
    \put(-278,-13.0){ $\bm\nearrow$ }
    \put(-296,65.0){ (a) }
    \put(-124,65.0){ (b) }
     \caption{\label{fig:graphene_lattice} 
(a) 2D graphene honeycomb lattice with a single impurity (turquoise sphere) at the center of one of the hexagonal cells.
(b) We show the distribution of induced charges on some of the lattice sites of pristine graphene as a function of plasmon energy (third dimension or Z-axis, i.e., direction of skewers pointing out of the plane above each lattice site).
The size and color of glyphs represent the magnitude and polarity of the induced charge, respectively.
We show the charge distribution only in the region of interest near the impurity site. Here we set the impurity potential $U_0=0$.  }
\end{figure}

In our case, we study the plasmonic excitations as a function of several parameters,
such as the sign (taking positive and negative sign of $U_0$),
strength (magnitude of $U_0$), and size (varying $\sigma$) of the impurity potential, and
the doping level (changing $\mu$) of the lattice.
To understand the interplay of these parameters in determining the plasmonic excitations we have to generate many data sets.
By varying just one of these parameters we introduce an extra dimension to the already 4D visualization problem, namely a fifth dimension.
A visualization tool that can handle this extra dimension is  desirable.
For example, in ParaView one can encode
the fifth variable (or dimension) as virtual time and selectively scan through a data set. 
The selective rendering of the data set, corresponding to different variables,
can be done from both a graphical user interface (GUI) and  script user interface (SUI) \cite{paraview_book, paraview_guide}. The
3D-rendered object can be displayed for multiple variable parameters  allowing a comparative study of the response of the system to these parameters.
Additionally, a quantitative analysis of the data set can be performed by using special filter functions already provided by the visualization tool or custom developed \cite{paraview_book, paraview_guide}.

For time efficient visualization, we used the server-client mode of parallel rendering in ParaView to analyze and visualize the data sets. 
We checked the scaling of the problem by recording the rendering time for different number of processors. 
We found that it scaled linearly for up to 16 processors (the maximum number we tested).
The results presented below were obtained by scanning through the data sets in the multidimensional parameter space.

\section{Plasmonic modes in graphene}

In Figure~\ref{fig:graphene_lattice}(a) we show the geometry of the graphene lattice used for our calculations.
The black dots specify the positions of carbon lattice sites.
The honeycomb lattice used in our calculations has $N=96$ sites. 
The  turquoise colored sphere, which sits at the center of a hexagonal cell,
represents the position of the single impurity.  All results presented here are for a single impurity on a 2D lattice with periodic boundary conditions.
The properties of the impurity are modeled by their effects on the on-site potentials
at neighboring carbon sites. 
Unless otherwise specified, the spatial extent of the impurity potential is restricted to the nearest neighbor carbon sites. 
Throughout this work, the energy is measured in units of the hopping energy, $t=2.7$ eV, 
between nearest neighbor carbon sites of pristine graphene.

\subsection{Pristine graphene}

To study the plasmonic excitations in pristine graphene, we set the impurity potential $U_0=0$ in our calculations.
In Fig.~\ref{fig:graphene_lattice}(b) we show the distribution of induced charges for a small
section of the graphene lattice as a function of energy. The Z-axis (here and in what follows)
 represents the energy $\omega$ of the plasmonic excitations.  We show the charge distribution only for a small
 section of the lattice, because later on this will be the region of interest for a spatially small-sized impurity potential.
  The color and size of the glyphs represents the polarity and the magnitude of the
induced charges, respectively. This visual image will be used as a reference image for a comparative study of the effects
 of a single impurity on nano-plasmonic excitations concentrated near an impurity.

\begin{figure}
  \centering
   \begin{tabular}{|c|c|}
   	\hline
  \includegraphics[width=3.5cm, height=3cm]{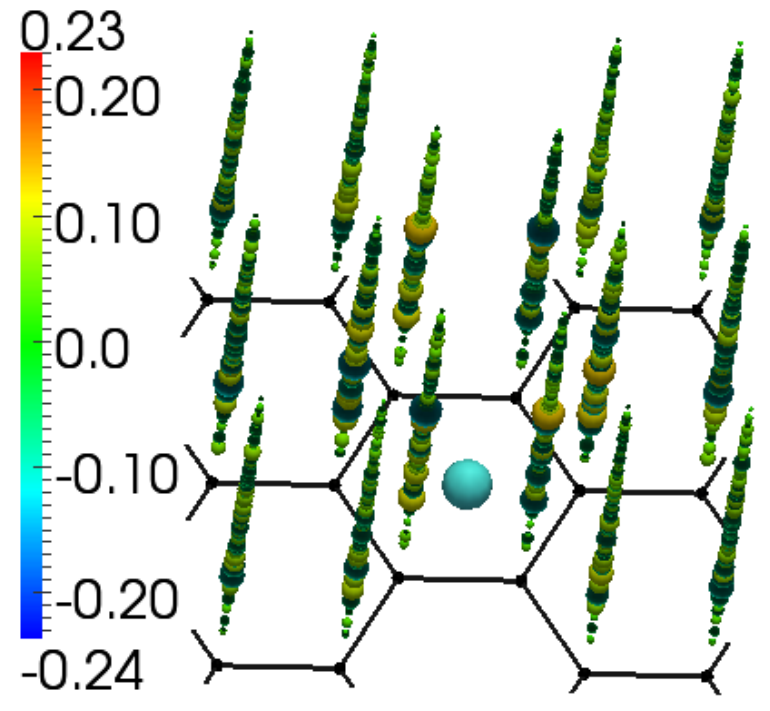} &
  \includegraphics[width=3.5cm, height=3cm]{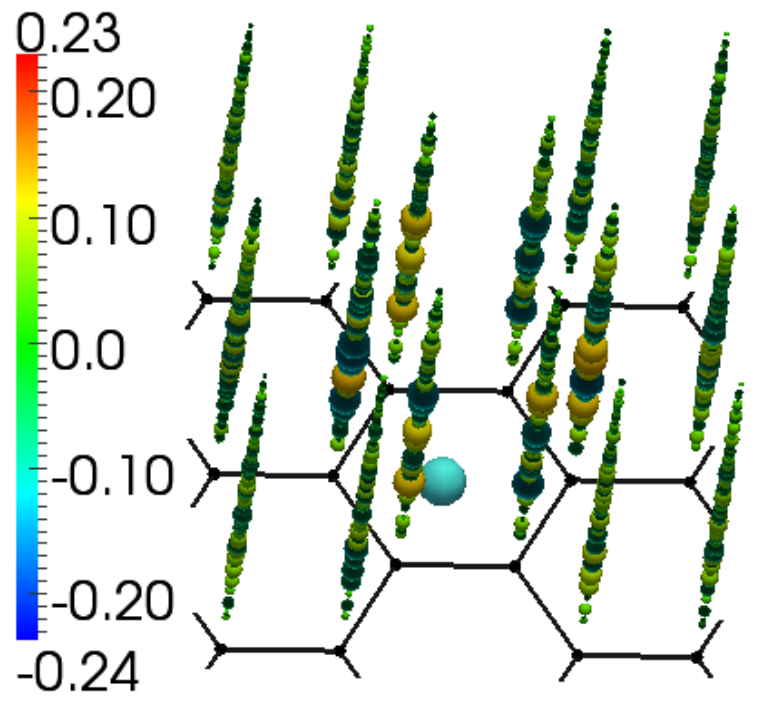} \\
  	\hline
  \includegraphics[width=3.5cm, height=3cm]{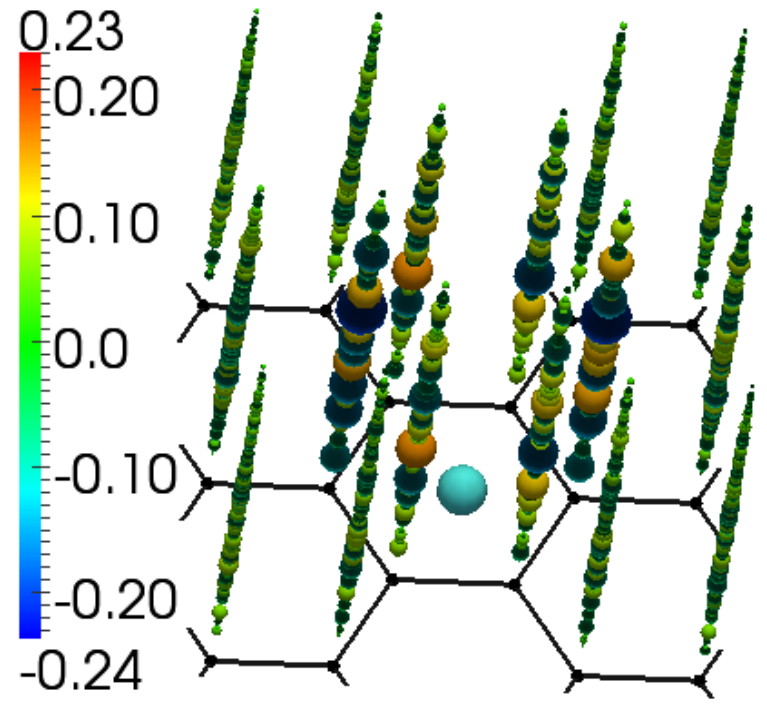} &
  \includegraphics[width=3.5cm, height=3cm]{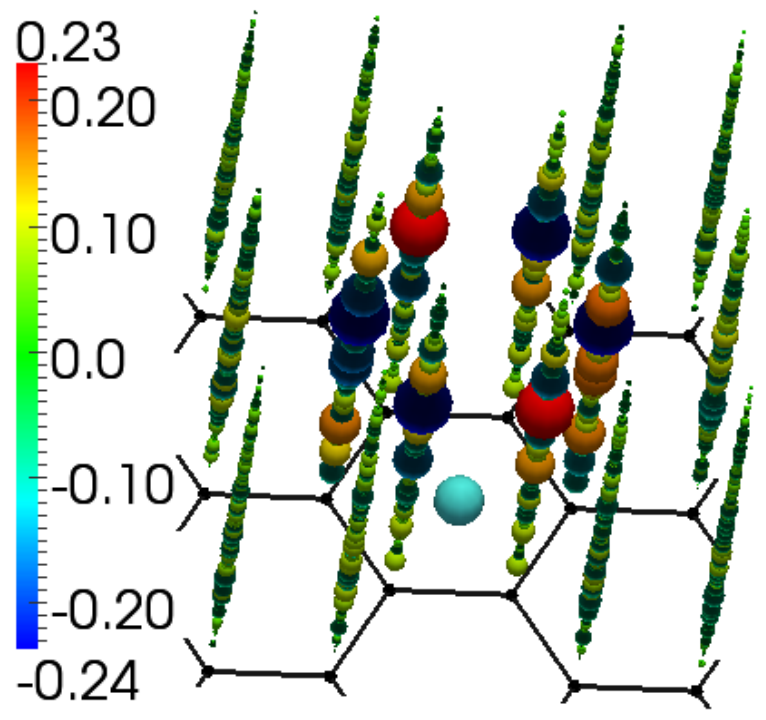} \\
  	\hline
  \end{tabular}
  \put(-190,84.0){ (a) }
     \put(-87,84){ (b) }
     \put(-190,-5){ (c) }
     \put(-87,-5){ (d) }
     \caption{\label{fig:positive_impurity} 
Distribution of the induced charges, on several lattice sites near the impurity,
as  a function of plasmon energy $\omega/t$.
Results are shown for $U_0/t=0.47$ (a), $0.79$ (b), $1.42$ (c), and $2.05$ (d),
where $t=2.7$ eV is the hopping parameter, which is used as the unit of energy.
The strength $U_0$ of the impurity drastically modifies the local response of the lattice.
The impurity strongly localizes the induced charges near the impurity site. }
\end{figure}

\subsection{Effect of positive and negative impurity potential at zero doping}

We study the effect of  a single impurity on the spatial distribution of the induced charges in the close vicinity of the impurity site for all plasmonic energies.
First we discuss the effect of an impurity with a repelling (positive) potential.
In Figure~\ref{fig:positive_impurity} we show results for the
impurity potential $U_0/t$=0.47, 0.79, 1.42, and 2.05, respectively.
 Comparing Fig.~\ref{fig:graphene_lattice}(b) with Fig.~\ref{fig:positive_impurity}(a),
one can see that the region closest to the impurity site shows the largest change in induced charge distribution. 
As the impurity potential is increased the change  becomes more pronounced, see
Figs.~\ref{fig:positive_impurity}(b)-(d), as seen in the increase in the size of the 3D sphere glyphs used to represent the magnitude of the induced charges.
More precisely, induced charges start increasingly to localize on neighboring sites of the impurity. 
These are the sites which are maximally affected by the impurity potential.

\begin{figure}
  \centering
   \begin{tabular}{|c|c|}
   	\hline
  \includegraphics[width=3.5cm, height=3cm]{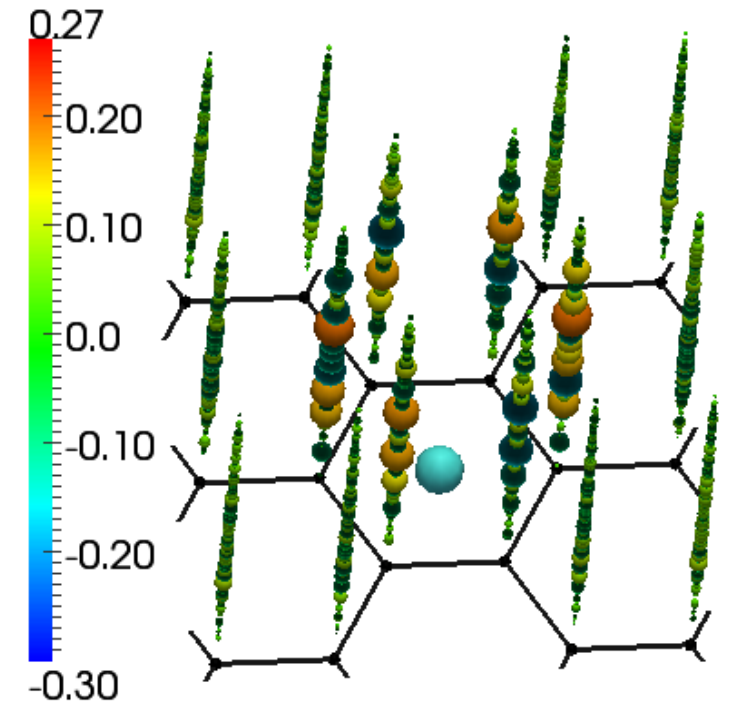} &
  \includegraphics[width=3.5cm, height=3cm]{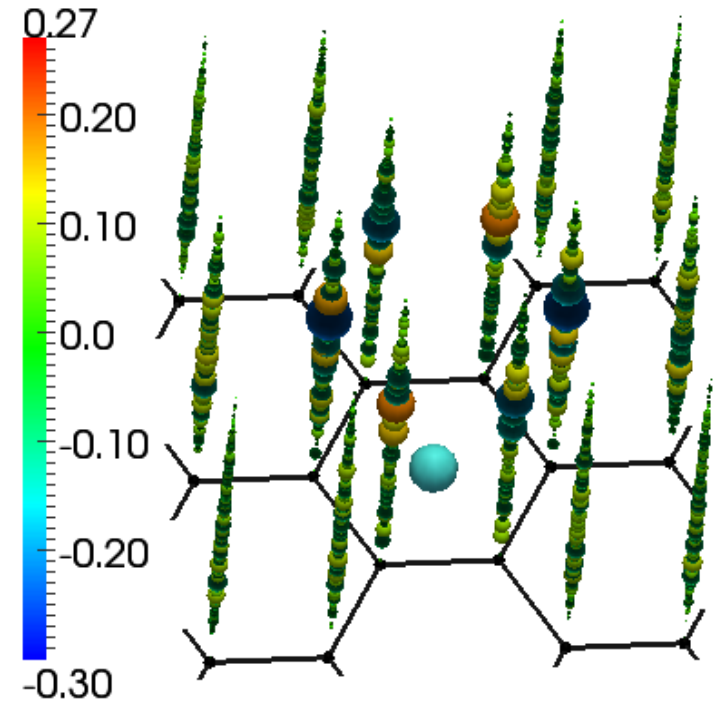} \\
  	\hline
  \includegraphics[width=3.5cm, height=3cm]{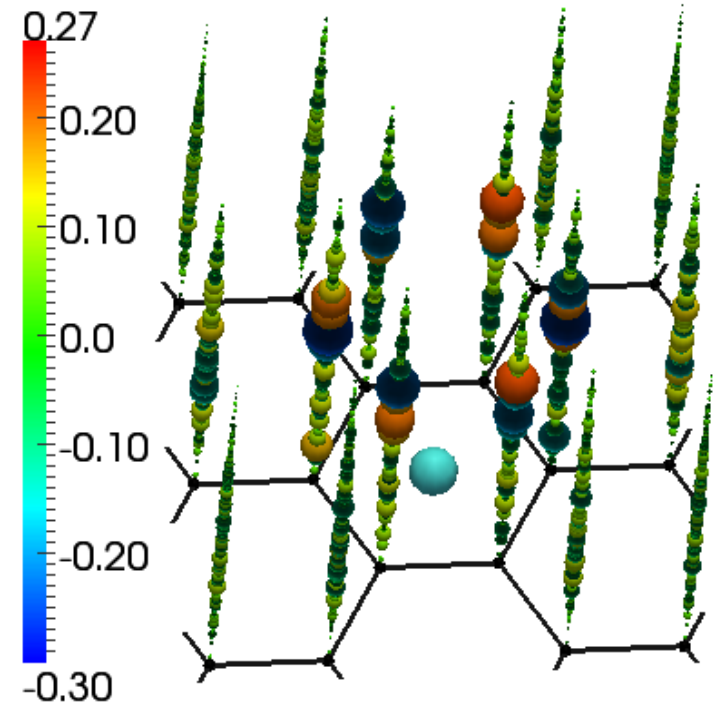} &
  \includegraphics[width=3.5cm, height=3cm]{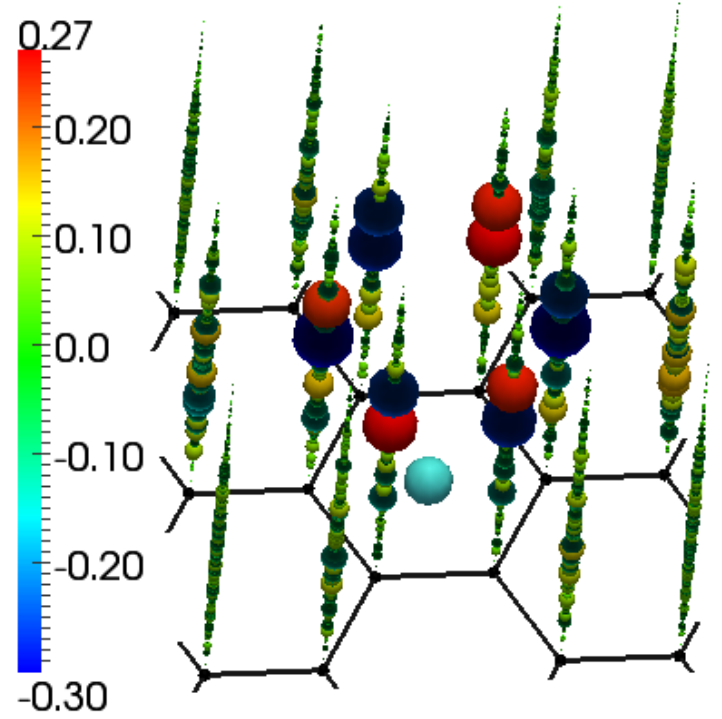} \\
  	\hline
  \end{tabular}
   \put(-192,84.0){ (a) }
     \put(-90,84){ (b) }
     \put(-192,-5){ (c) }
     \put(-90,-5){ (d) }
      \put(-182.5,24.0){ $\uparrow$ }
        \put(-79.5,23.50){ $\uparrow$ }
        \put(-183.5,-65.0){ $\uparrow$ }
        \put(-79.5,-65.0){ $\uparrow$ }
     \caption{\label{fig:size_effect} 
Distribution of induced charges, on several lattice sites near the impurity, as  a function of energy $\omega$.
Results are shown for  increasing the spatial extent of the impurity potential by changing its Gaussian width 
$\sigma/a_0=1.0$ (a),  $1.2$ (b), $1.3$ (c), and $1.4$ (d), where $a_0$ is the lattice constant. We fixed $U_0=2.0t$.
Increasing the impurity size localizes more charges close to the impurity.  It also enhances the weight of the induced charge on the next-nearest-neighbor lattice sites pointed out in the figure by arrows. }
\end{figure}
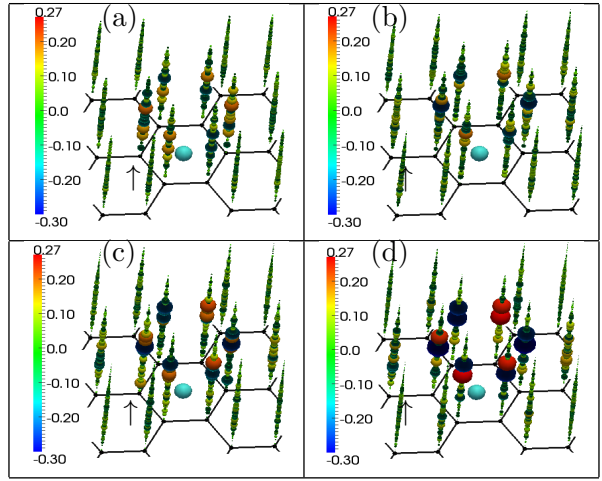

We also study the effect of an attractive impurity potential on the induced charge distribution for  $U_0/t=-0.47$,$-0.79$, $-1.42$, and $-2.05$. Similar to the case of a positive impurity potential,  the most pronounced change
occurs on lattice sites in close proximity to the impurity. As the impurity potential is increased more and more induced charge is localized at these sites to more efficiently screen the impurity.

For given magnitude of the impurity potential, we do not see a visual difference in the induced charge distribution between the positive and negative impurity. To be more accurate about this visual inference, we quantitatively compare the plasmonic excitations at all energies 
as well as their corresponding spatial distribution of induced charges for positive and negative impurities (see Sec.~5 for more information on how we extract this information). Indeed, we confirm that our calculations satisfy the sum rule for all plasmonic excitations and that the corresponding induced charge distributions are identical for attractive and repelling impurity potential at half filling.

The exact same response of the graphene lattice for both positive and negative single impurity at zero doping is related to the fundamental particle-hole symmetry of graphene. The tight-binding calculation shows that the valence and conduction band meet at two nonequivalent points in the Brillouin zone, which are called Dirac points. At zero doping the valence band is completely filled and the conduction band is empty (half filling). The Fermi level lies at the Dirac point so valence and conductance bands are symmetric with respect to the Fermi level (particle-hole symmetric). The same response of the lattice for positive and negative impurity potential originates from the particle-hole symmetric electronic band structure. We use this observation as a validity check for the numerical method implemented to calculate the plasmonic response in graphene. Later on we will see that once we break this symmetry, for example, by doping the system 
away from half filling ($\mu\neq 0$), the response of the system will be different for positive and negative single impurity.

 \begin{figure}
  \centering
   \begin{tabular}{|c|c|}
   	\hline
  \includegraphics[width=3.5cm, height=3cm]{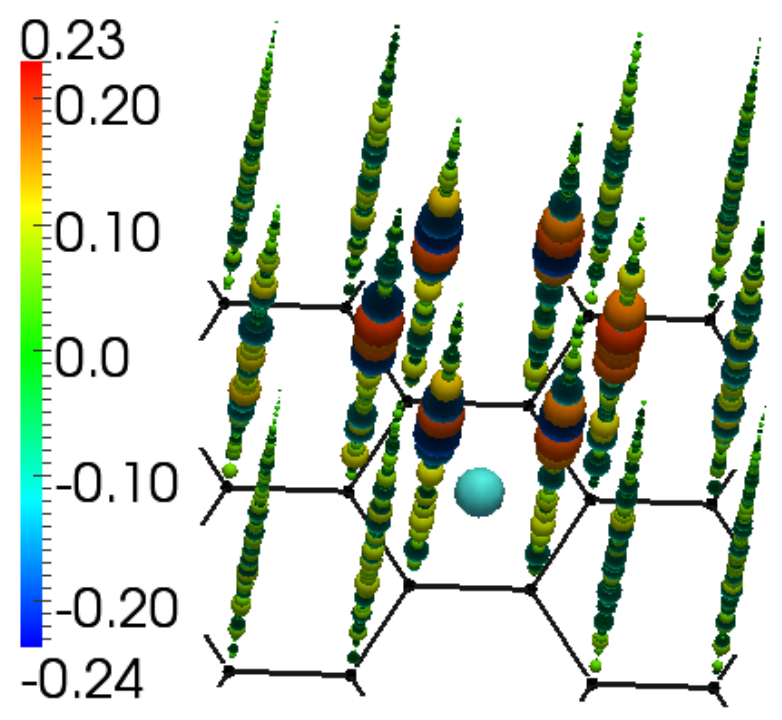} &
  \includegraphics[width=3.5cm, height=3cm]{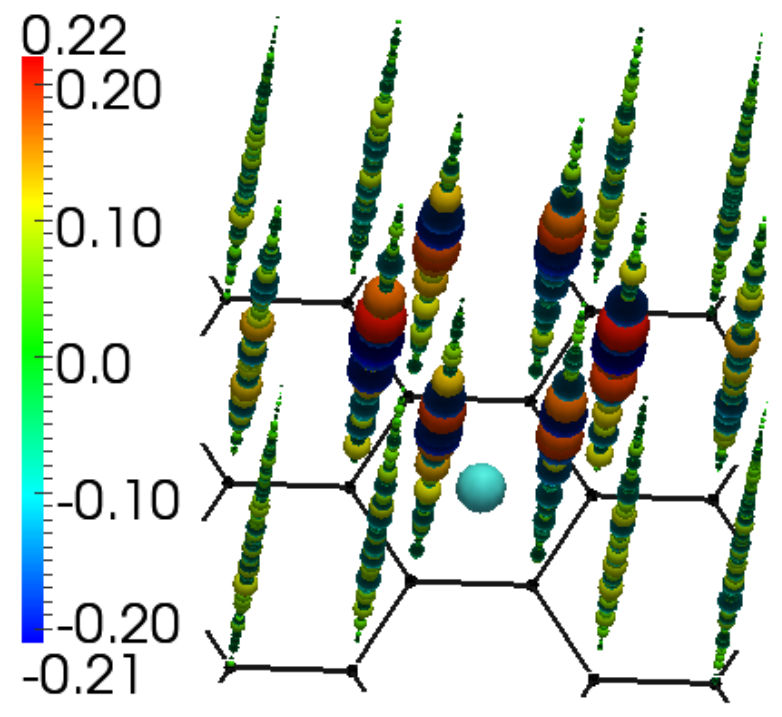} \\
  	\hline
  \includegraphics[width=3.5cm, height=3cm]{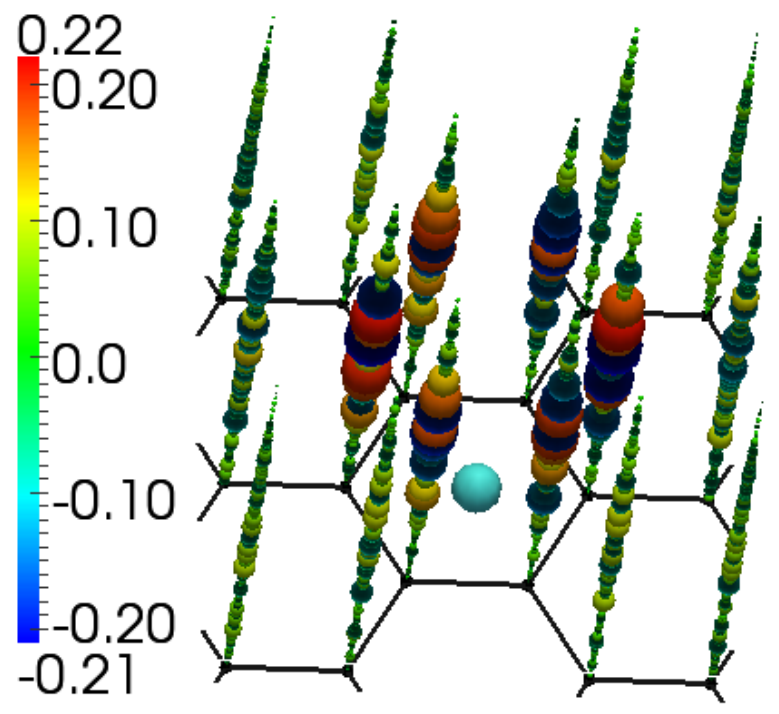} &
  \includegraphics[width=3.5cm, height=3cm]{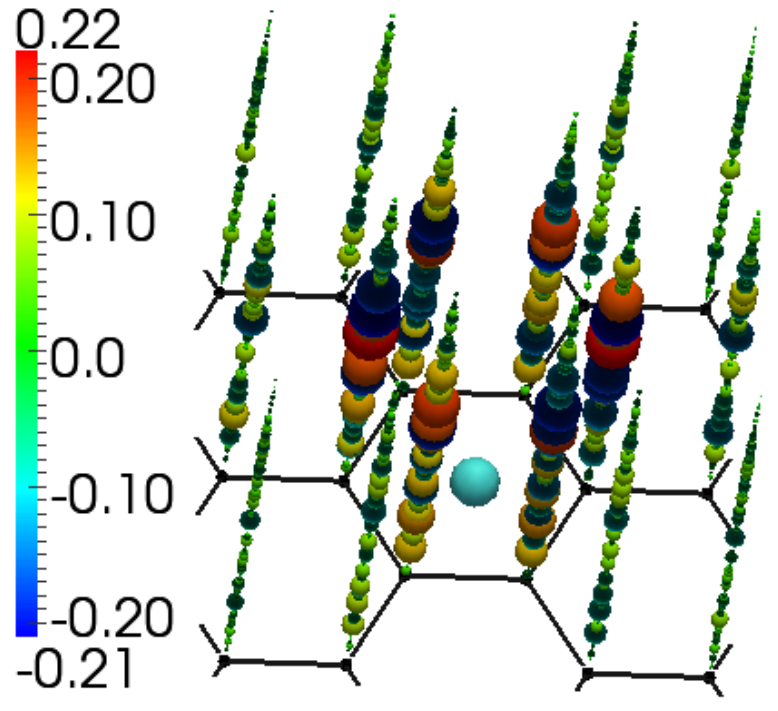} \\
  	\hline
  \end{tabular}
   \put(-190,84.0){ (a) }
     \put(-85,84){ (b) }
     \put(-190,-5){ (c) }
     \put(-85,-5){ (d) }
     \caption{\label{fig:electron_doping} 
Distribution of induced charges in graphene with positive impurity potential $U_0=3.0t$ and
electron doping $\mu/t=0.0$ (a), $0.75$ (b), $1.5$ (c), and $2.5$ (d).
As the electron doping is increased the localization of the charges at sites close to the impurity site are enhanced,
 while it is decreased for lattice sites farther away. The number of localized plasmons (nano-plasmons) is increased as
seen in the increasing number of bigger 3D sphere glyphs (charges) along the energy axis. In other words, screening 
by conduction electrons is increased in the vicinity of the single impurity.
  }
\end{figure}

\subsection{The effect of impurity size}

A spatially extended impurity is modeled by taking into account the effects of the impurity on lattice sites beyond nearest neighbor carbon sites. This is mathematically achieved by relaxing the decay of the impurity  potential over distance by increasing the magnitude of $\sigma/a_0$ from $1.0$ to $1.4$, where $a_0$ is the lattice constant. Here we fix the potential to $U_0=2.0t$.

In Figures~\ref{fig:size_effect}(a)-(d), we show the effects of increased impurity size on the induced  charge distributions. 
Comparing Fig.~\ref{fig:size_effect}(a) with Figs.~\ref{fig:size_effect}(b)-(d), one can see that as the size of the impurity potential is increased the localization of the induced charges in close vicinity of the impurity is increased (notice the increase in the range of the induced charge in the color legend of Fig.~\ref{fig:size_effect} compared to that of Fig.~\ref{fig:positive_impurity}). 
Moreover, we see that it also enhances the weight of the induced charges on the next-nearest-neighbor carbon sites (compared to the magnitude of the charges at the lattice site where the arrow points to). 
Here too, we find that more and more induced charges are localized closer to the impurity with an increase of the extent of the impurity.


\subsection{The effect of positive and negative impurity potential at finite doping}

We model the effect of doping by taking a nonzero value for the chemical potential $\mu$
in Eq.~(\ref{eqn:hamiltonian}). For electron doping the chemical potential moves above
the Dirac point and lies in the conduction band, i.e., the chemical potential
is positive.  In Figures~\ref{fig:electron_doping}(a)-(d) we show the effects of the impurity on the induced charge distributions for doping levels $\mu/t=0.0$, $0.75$, $1.5$, and $2.5$, respectively. The strength of the impurity potential is $U_0=3t$. 
Comparing Fig.~\ref{fig:electron_doping}(a) with Figs.~\ref{fig:electron_doping}(b)-(d), one can see that as the level of electron doping is increased the localization of the induced charges at sites close to the impurity are enhanced. The magnitude of the induced charges on the remaining lattice sites is consequently suppressed. Moreover, the number of localized plasmon (nano-plasmons) increases as seen in the increased number of bigger 3D sphere glyphs along the energy axis. This effect is most clearly seen in Fig.~\ref{fig:electron_doping}.

The origin for the increase in the number of nano-plasmons
 (in the presence of a positively charged impurity), due to  electron doping, comes from 
a change in the electronic and impurity states of graphene \cite{rodrigo2010}.
 The impurity states lie at the band edge and electron doping brings the top of the
filled graphene electronic states closer to the impurity states.
This leads to an increase in the number of localized excitations near the impurity site.


\begin{figure}
  \centering
   \begin{tabular}{|c|c|}
   	\hline
  \includegraphics[width=3.5cm, height=3cm]{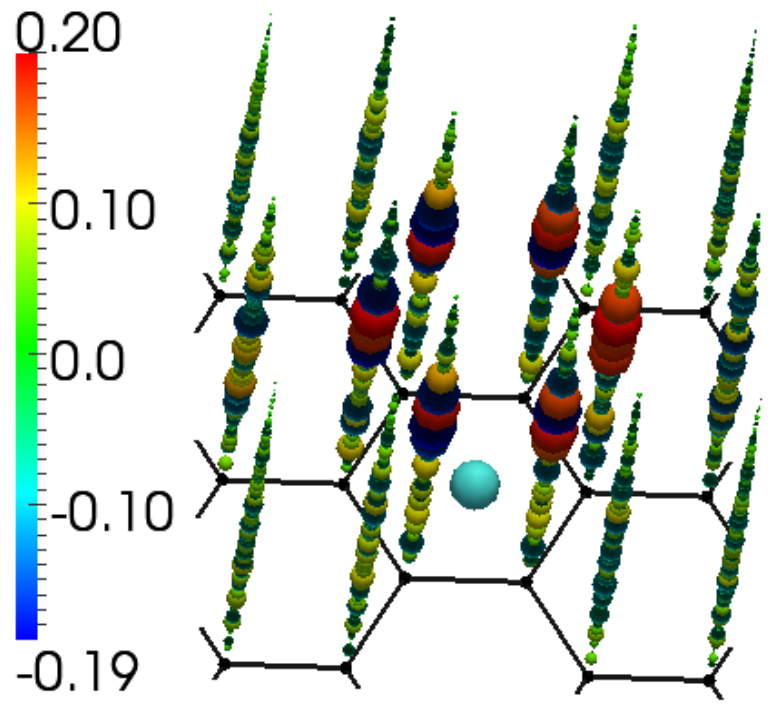} &
  \includegraphics[width=3.5cm, height=3cm]{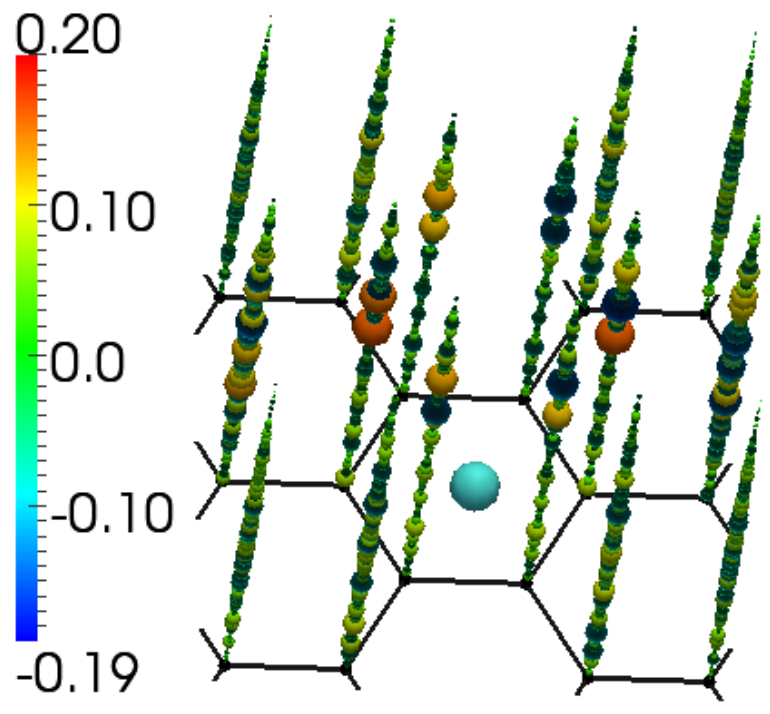} \\
  	\hline
  \includegraphics[width=3.5cm, height=3cm]{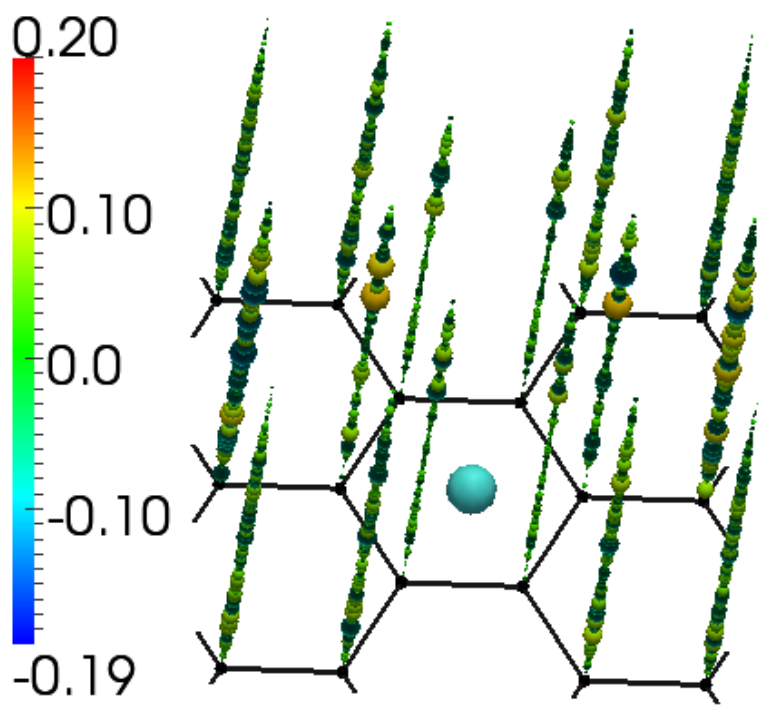} &
  \includegraphics[width=3.5cm, height=3cm]{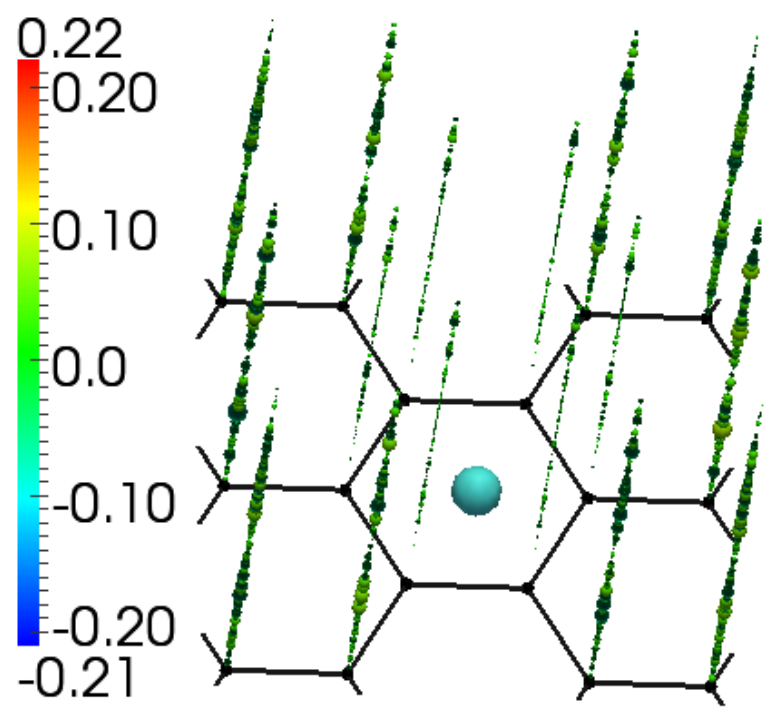} \\
  	\hline
  \end{tabular}
  \put(-192,84.0){ (a) }
     \put(-87,84){ (b) }
     \put(-192,-5){ (c) }
     \put(-87,-5){ (d) }
     \caption{\label{fig:hole_doping} 
Distribution of induced charges in graphene with positive impurity potential $U_0=3.0t$ and
hole doping levels of $\mu/t=0.0$ (a), $-0.75$ (b), $-1.5$ (c), and $-2.5$ (d).  
As the hole doping is increased the localization of charges at sites close to the impurity is reduced. }
\end{figure}

When the system is hole doped, the chemical potential moves below the Dirac point and lies in the valence band. The effect is modeled by taking a negative value for the chemical potential in Eq.~(\ref{eqn:hamiltonian}). 
In Figure~\ref{fig:hole_doping}(a)-(d) we show the effects of the impurity on the charge distributions for doping levels 
$\mu/t=0.0$, $-0.75$, $-1.5$, and $-2.5$, respectively. 
The strength of the impurity potential is fixed at $U_0=3.0t$. Comparing Fig.~\ref{fig:hole_doping}(a) with the Figs.~\ref{fig:hole_doping}(b)-(d), one can see that as the hole doping is increased the localization of  charges in close vicinity of the impurity is suppressed. The suppression of localization of nano-plasmons is caused by the increased separation between the top of the filled states of graphene and the impurity states. The different nano-plasmonic responses of the graphene lattice due to electron and hole doping originates from the breaking of particle hole-symmetry with doping away from 
half filling.


\begin{figure}
  \centering
   \begin{tabular}{|c|c|}
	\hline
\includegraphics[width=4.1cm, height=5.0cm,angle=0]{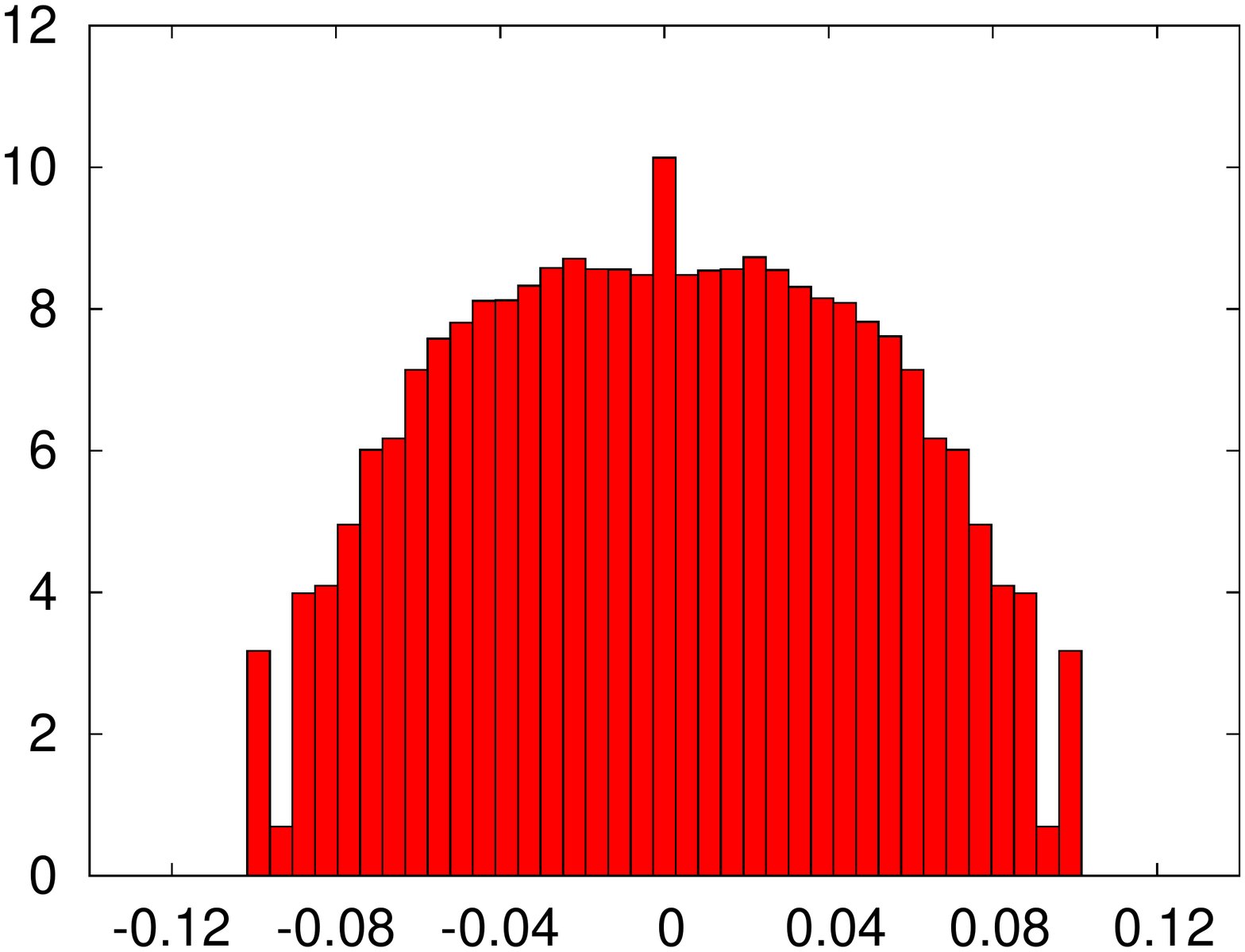} &
\includegraphics[width=4.1cm, height=5.0cm,angle=0]{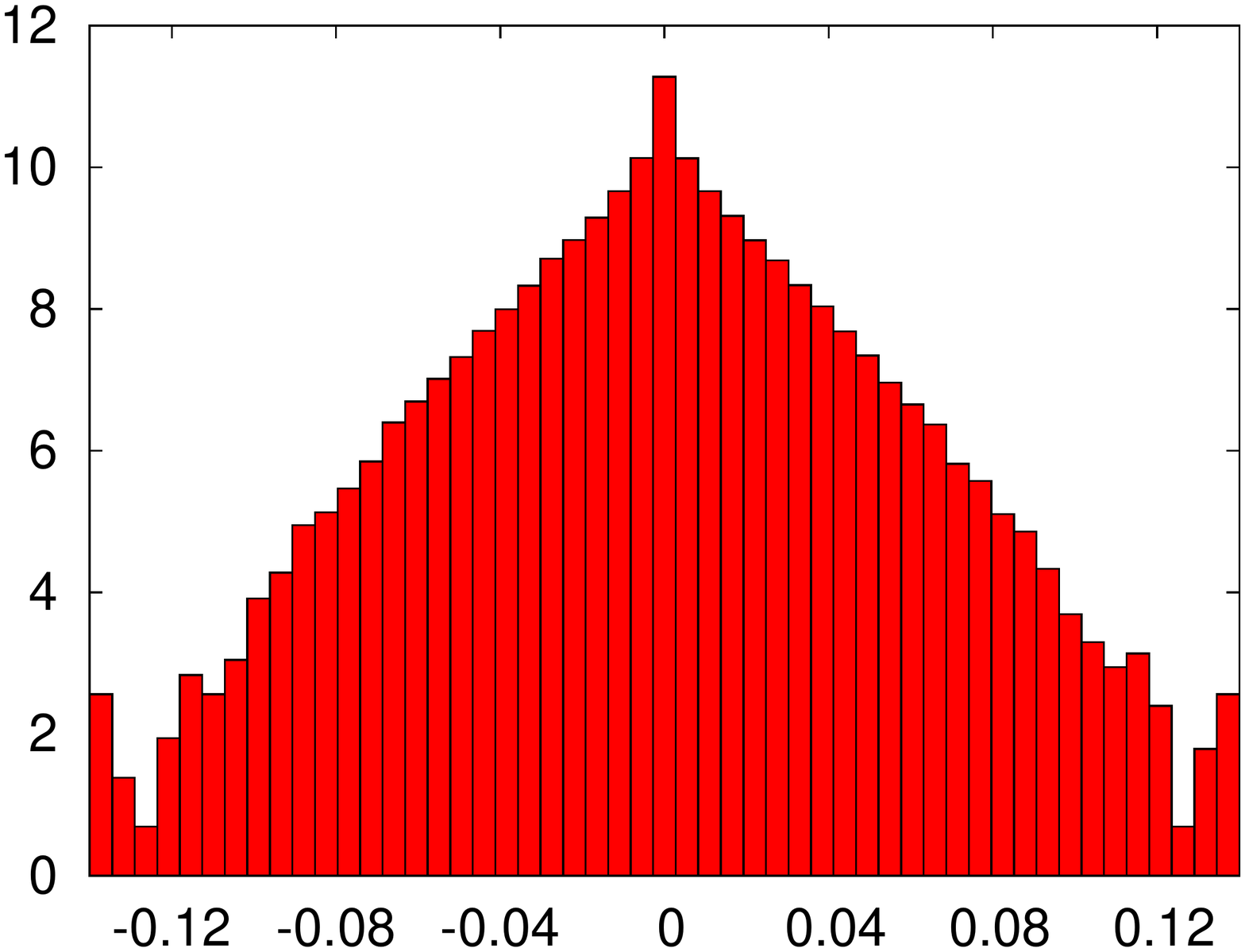} \\
	\hline
  \end{tabular}
      \put(-260,65){ (a) }
     \put(-132,65){ (b) }
      \put(-260,-5){
      \rotatebox{90}{ $log(N)$}}
     \put(-135,-5){ {
      \rotatebox{90}{ $log(N)$}}}
      \put(-193,-65){ $\Delta \rho$}
     \put(-65,-65){ $\Delta \rho$}
     \caption{\label{fig:histogram} 
Probability function of induced charges over the entire 2D lattice
in (a) pristine graphene  and (b) graphene with a single impurity with a positive potential $U_0=1.0t$.
The distribution function of  charges is symmetric with respect to the origin. The range of the
magnitude of induced charges is increased in impure graphene. }
\end{figure}

To summarize, we studied  nano-plasmonic excitations with a fixed negative impurity potential but changing the chemical potential. We found that in the hole (electron) doped case the negative impurity potential leads to an increase (decrease) of the localization of induced charges near the impurity. Thus a positive impurity potential for electron doping behaves similarly as a negative impurity potential for hole doping. Both of them enhance the localization around the impurity. 
In contrast, a positive impurity potential for hole doping and a negative impurity potential for electron doping suppress localization of induced charges around the impurity.  These results are listed in Table~\ref{table:summary}. It is worth to notice that for a given magnitude of the chemical potential and impurity potential, flipping the sign of both of them gives identical nano-plasmonic response in energy and real space.

\begin{figure}
  \centering
   \begin{tabular}{|c|c|}
	    \hline
  \includegraphics[width=4.1cm, height=4.50cm]{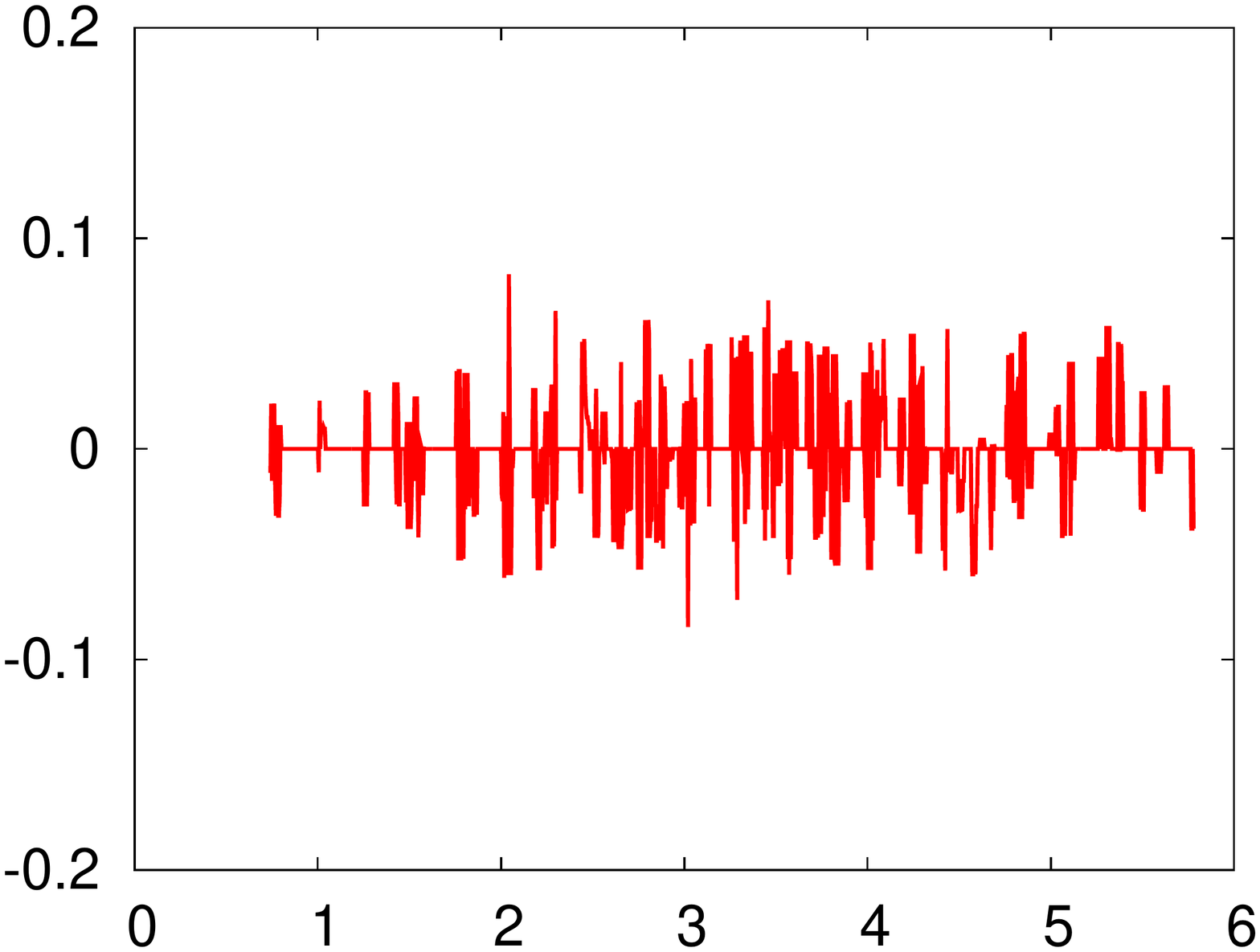} &
  \includegraphics[width=4.1cm, height=4.50cm]{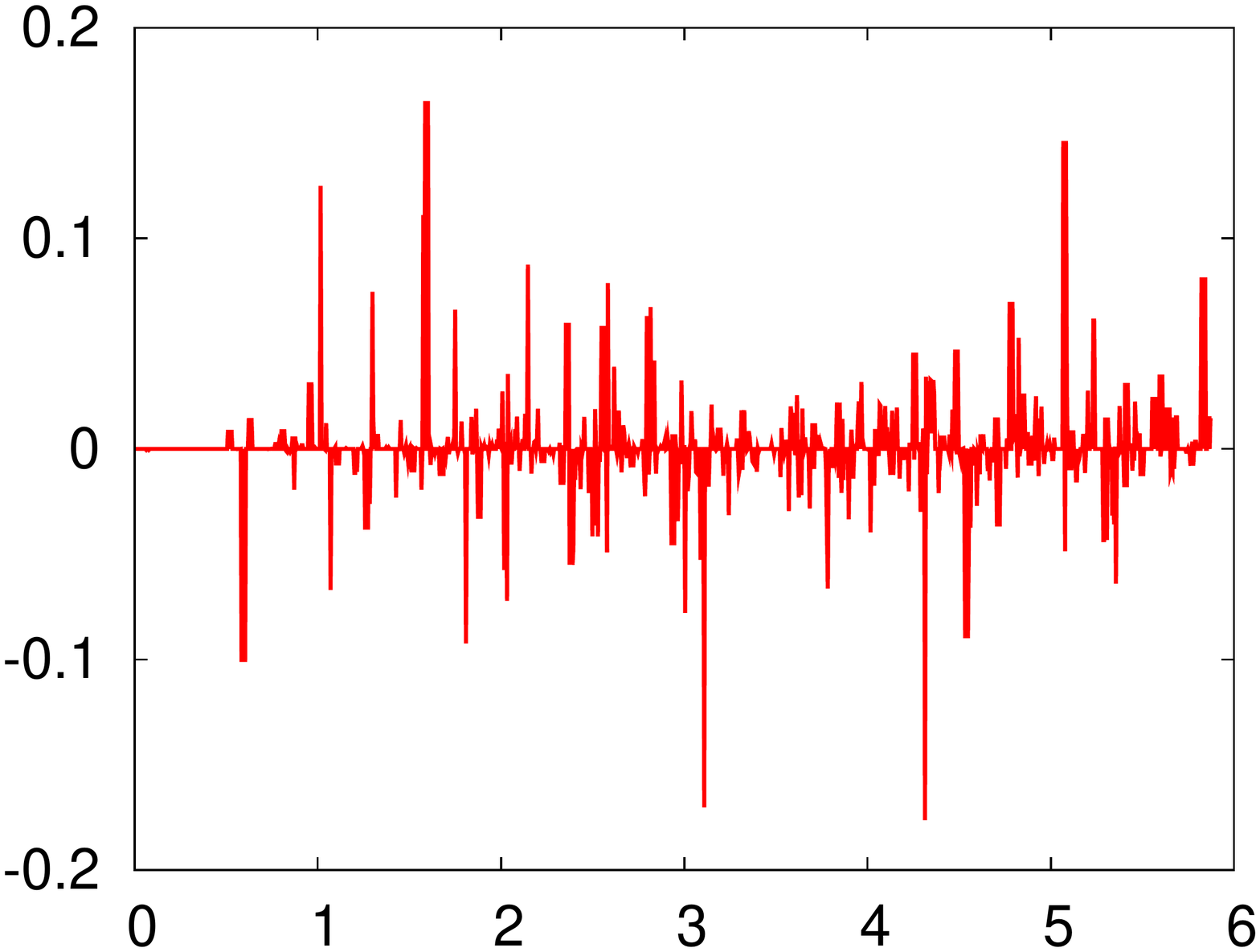} \\
	    \hline
  \end{tabular}
      \put(-260,58){ (a) }
     \put(-132,58){ (b) }
      \put(-260,8.0){ $\Delta \rho$}
       \put(-132,8.0){ $\Delta \rho$}
      \put(-200,-57.0){ $\omega/t$}
       \put(-70,-57.0){ $\omega/t$}
     \caption{\label{fig:probe_line} 
Induced charges in (a) pristine and (b) impure graphene as a function
of energy at a fixed nearest-neighbor lattice site from the impurity, 
marked by the arrow in Fig.~\ref{fig:graphene_lattice}(a).
The result is for a positive impurity potential $U_0=2t$. We see a larger range of  
induced charges in  impure graphene.}
\end{figure}

\begin{table}[ht]
\centering
\caption{Dependence of nano-plasmons in graphene on the sign of the impurity potential and doping (chemical potential).}
\begin{tabular}{|c|c|c|}
	\hline
impurity potential &  charge doping &  plasmonic localization \\
	\hline
positive & electron & enhanced  \\
negative & hole & enhanced  \\
positive & hole & suppressed  \\
negative & electron & suppressed  \\
	\hline
\end{tabular}
\label{table:summary}
\end{table}


\begin{figure*}
  \centering
   \begin{tabular}{|c|c|c|c|}
  	 \hline
  \includegraphics[width=3cm, height=3cm]{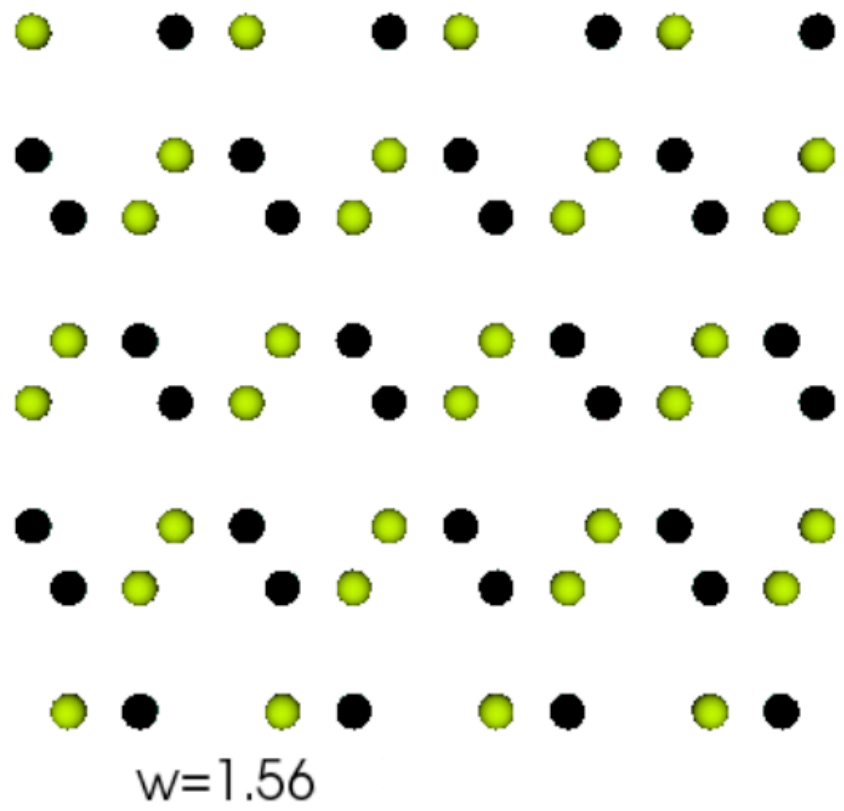} &
  \includegraphics[width=3cm, height=3cm]{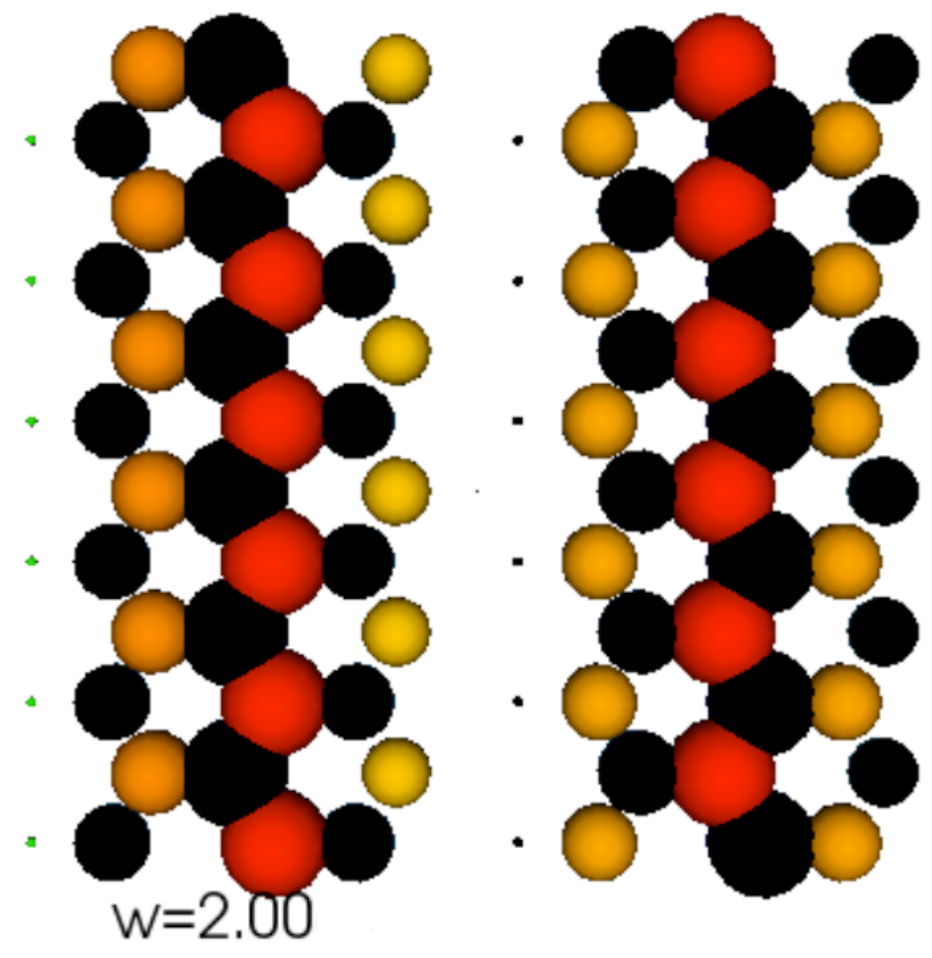} &
  \includegraphics[width=3cm, height=3cm]{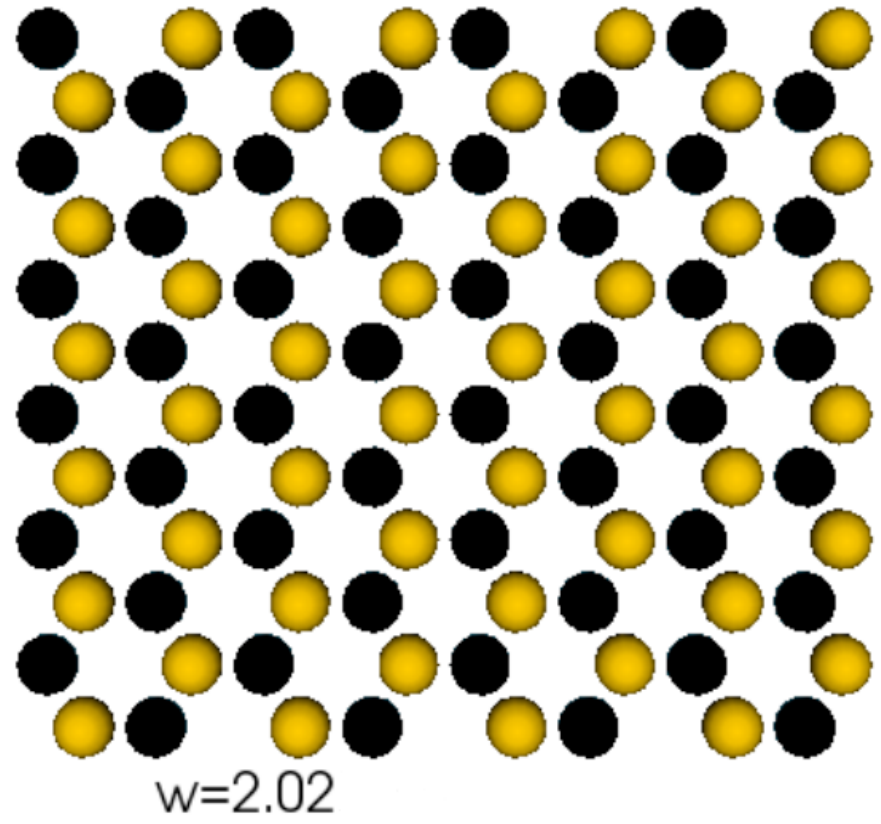} &
  \includegraphics[width=3cm, height=3cm]{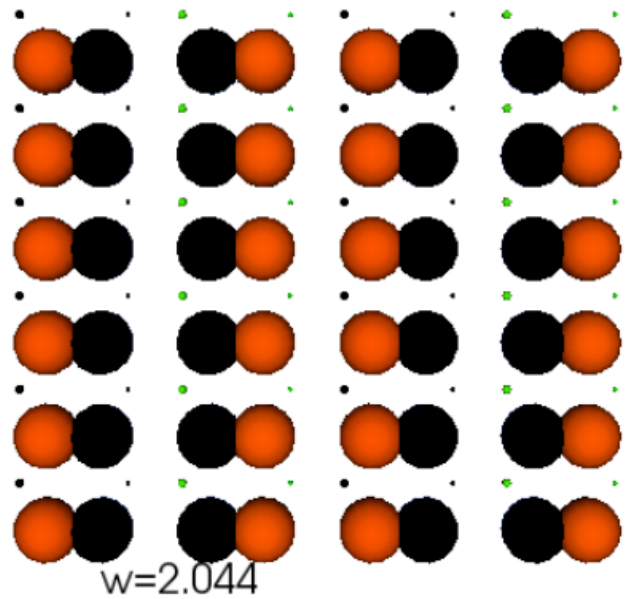} \\
 	 \hline
  \includegraphics[width=3cm, height=3cm]{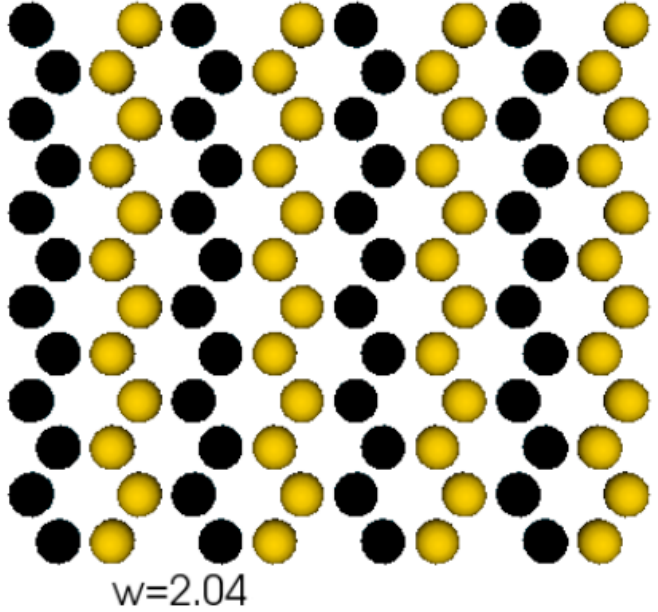} &
  \includegraphics[width=3cm, height=3cm]{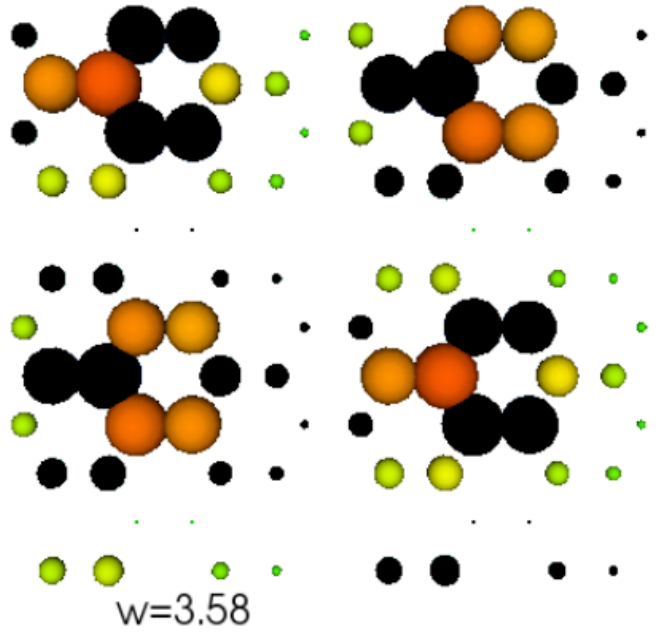} &
   \includegraphics[width=3cm, height=3cm]{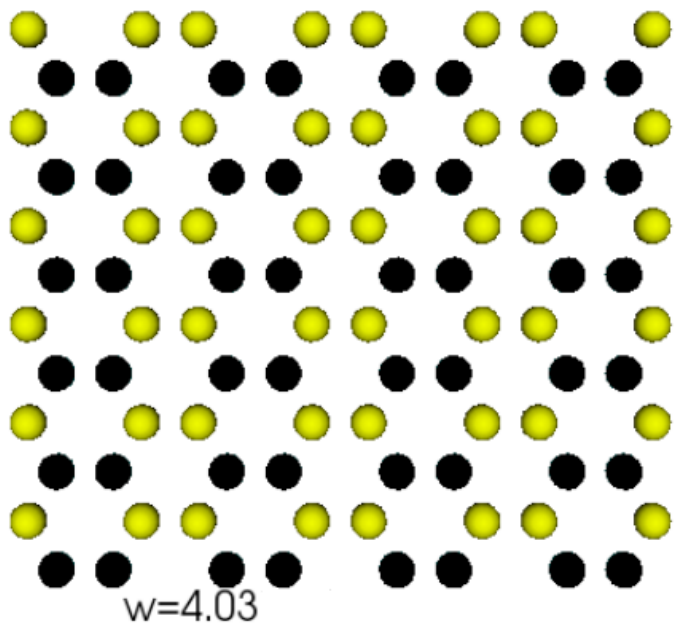} &
  \includegraphics[width=3cm, height=3cm]{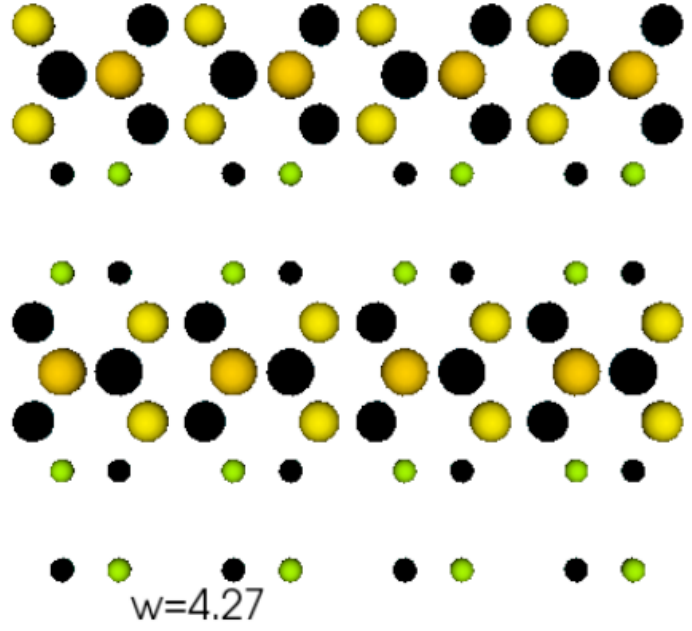} \\
  	\hline
  \end{tabular}
     \caption{\label{fig:plane_cut_U=0} 
The spatial distribution of the induced charges on every lattice site at given energy for
pristine graphene. Only a few selected modes are shown. The  energy ${\rm w}=\omega/t$ of plasmon labels the panels.}
	\end{figure*}

  \begin{figure*}
  \centering
   \begin{tabular}{|c|c|c|c|}
   	\hline
  \includegraphics[width=3cm, height=3cm]{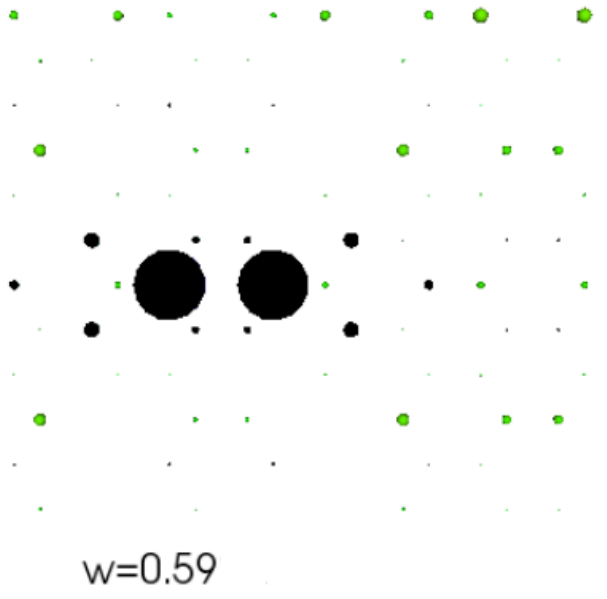} &
  \includegraphics[width=3cm, height=3cm]{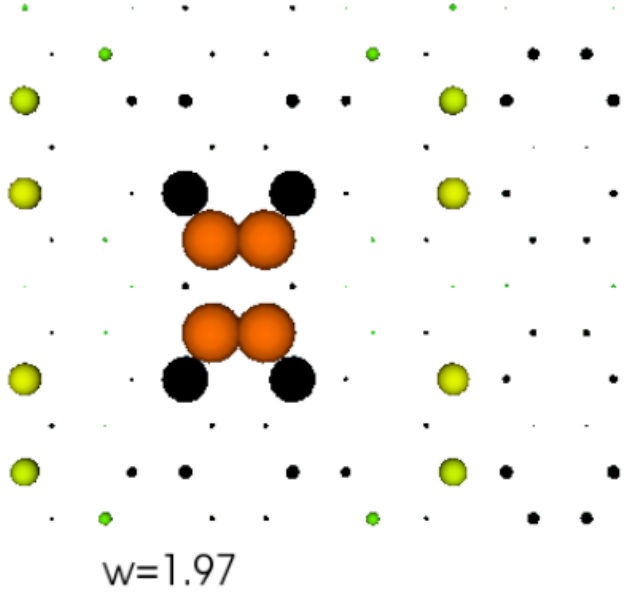} &
  \includegraphics[width=3cm, height=3cm]{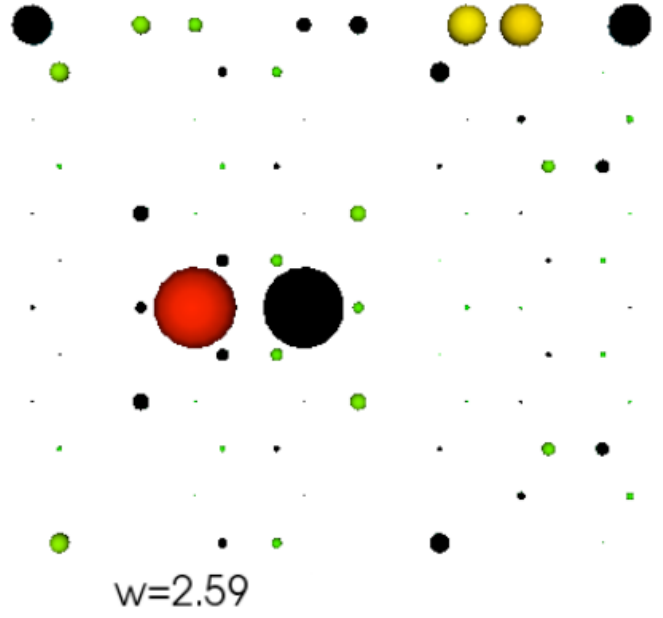} &
  \includegraphics[width=3cm, height=3cm]{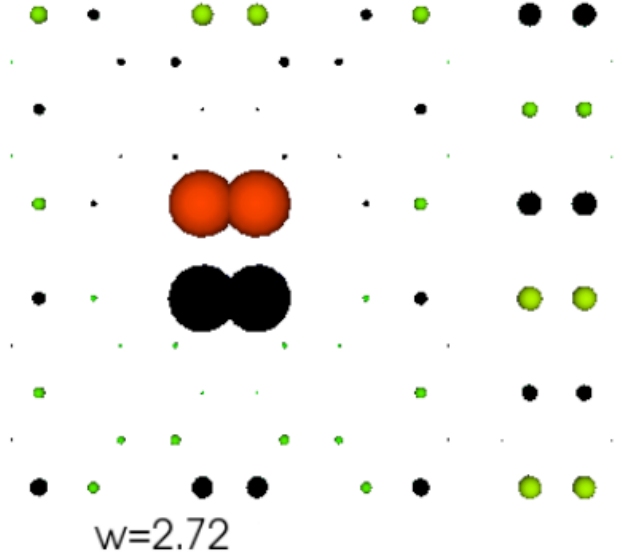} \\
  	\hline
  \includegraphics[width=3cm, height=3cm]{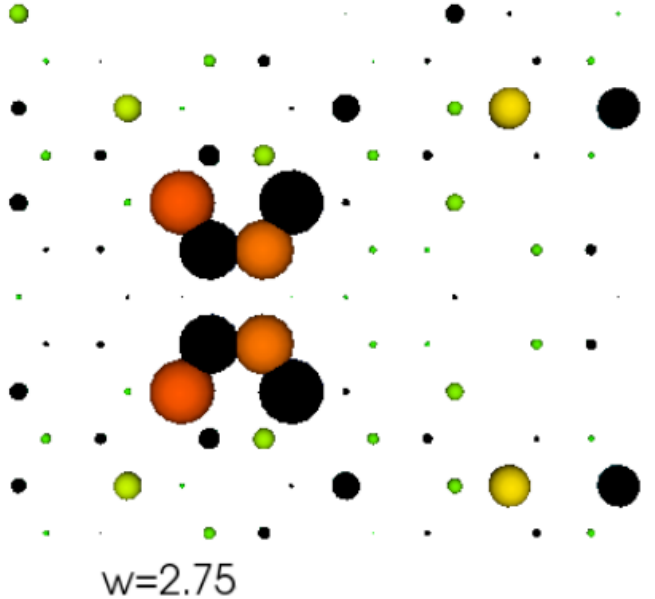} &
  \includegraphics[width=3cm, height=3cm]{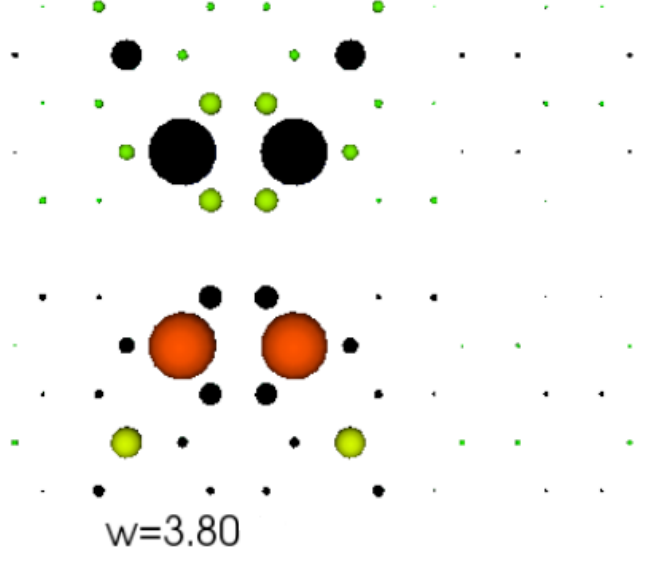} &
   \includegraphics[width=3cm, height=3cm]{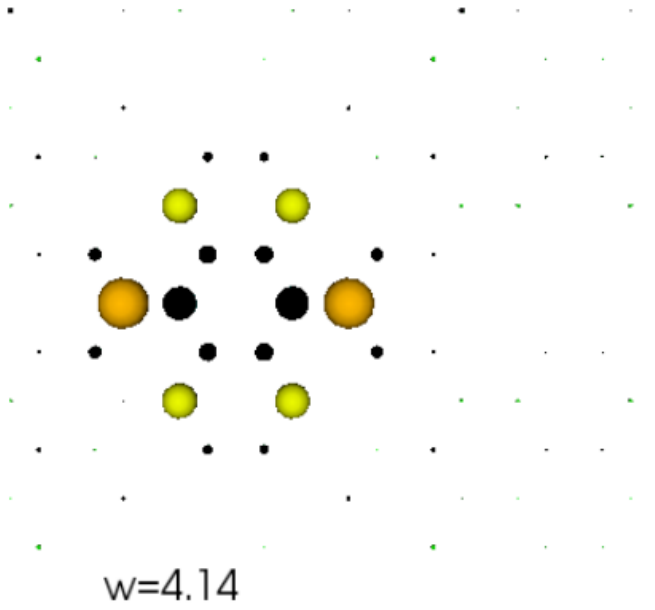} &
  \includegraphics[width=3cm, height=3cm]{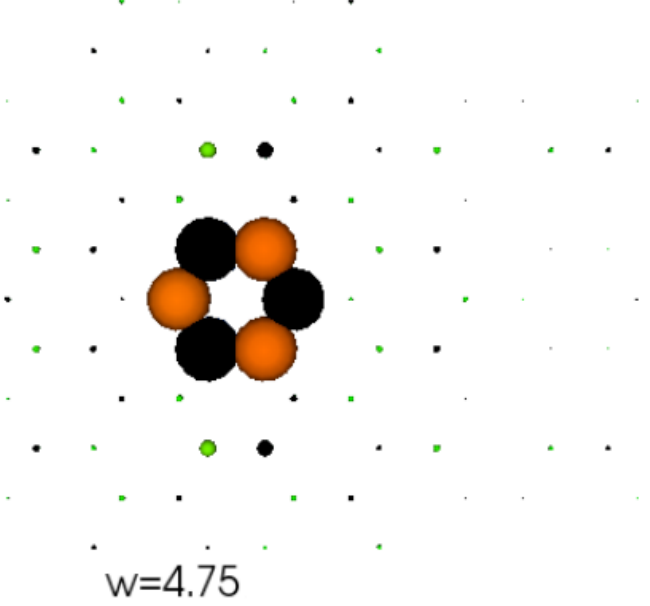} \\
  	\hline
  \end{tabular}
     \caption{\label{fig:plane_cut_U=2} 
The spatial distribution of the induced charges on
individual lattice sites at given energy near the impurity on graphene lattice.
Only a few selected nano-plasmons are shown. The  energy ${\rm w}=\omega/t$ of plasmon labels the panels. 
The localization of plasmons occurs at the nano-meter scale, hence the term {\it nano-plasmons}.  }
     \end{figure*}


\section{Quantitative analysis}

Using an advanced visualization tool like ParaView one can extract in addition to the {\it obvious}
visual cues \cite{Keller} also quantitative information from the data set of interest.
We demonstrate this capability by first plotting a histogram of the distribution of induced charges over all
lattice sites and over all energies. The histogram is generated by applying the built-in `Histogram' filter
in ParaView \cite{paraview_book, paraview_guide} on a data set with impurity potential $U_0=1.0t$. This information may appear trivial, but is of interest to verify the
conservation of total charge in our calculations, i.e., the sum of all induced charges must vanish and 
must be equally
distributed between positive and negative oscillations.
Since pristine graphene does not have localized plasmons, whereas  impure graphene does,
it is important to understand the rearrangement of induced charges.
We can check for charge conservation and induced charge balance for two completely different cases, namely, localization vs.\ delocalization of modes.
The symmetric distribution of the induced charges as seen in the histogram or probability function of finding a specific induced 
charge in Fig.~\ref{fig:histogram} 
implies that the total induced charge is zero. 
We verified the charge neutrality by calculating the sum of the  induced charges.
 
In both cases, pristine and impure graphene, the number of zero-induced charges is significantly larger
 than the number of nonzero induced charges, i.e, lattice sites that are unaffected. 
However the number of zero-induced charges is slightly bigger in impure
graphene than in pristine graphene. The magnitude of induced charges
decrease exponentially in impure graphene, while it has a Gaussian distribution in pristine graphene. 
The range of the magnitude of induced charges is slightly bigger for impure graphene.
This implies that a relatively small number of plasmons acquires a bigger magnitude of induced charges.
The larger number of zero-induced charges, as well as a wider range in the distribution of charges, and the exponential decrease  
are consequences of localized plasmonic excitations.

Next, we monitor the induced charge at a single site in the graphene lattice as a function of energy.
The calculation is performed with or without the presence of impurity. For the impurity case the impurity potential is fixed at $U_0=2.0t$.
We plot this information by using the built-in `Plot Over Line' filter \cite{paraview_book, paraview_guide}. 
It gives interpolated values of the variable along an arbitrary user-specified line.
The result is shown in Fig.~\ref{fig:probe_line}  for a particular nearest-neighbor site of the impurity, marked by
the arrow in Fig.~\ref{fig:graphene_lattice}(a). 
One also sees that the range
of induced charges are bigger for the impure graphene.  
In addition, comparing Fig.~\ref{fig:histogram}(b) and Fig.~\ref{fig:probe_line}(b)
one can see that the range of induced charges in impure graphene increases with increasing impurity potential
from $U_0=1.0t$ to $2.0t$.

Finally, we extracted the energy-resolved spatial distribution of induced charges
for pristine and impure graphene. As mentioned above,
there are about 2300 plasmonic modes for a lattice with $N=96$ sites.
It is very time consuming to extract every individual plasmonic excitation manually.
However, this can be automated in ParaView by using the SUI (script user interface). We developed a simple
{\it python} script, which  controls the rendering in ParaView,
to extract the distribution of the induced charges for all energies.
It is worth mentioning that using a single-processor computer it is impractical to extract the necessary
information in real time  (it takes several hours to resolve the spatial distribution for all energies).
This computationally intensive problem can be overcome by running the SUI 
in parallel rendering mode.
Since there are many unique plasmonic modes (characterized by the unique induced charge distributions),
 it is impossible to report all of them.
So we show only a few selected plasmonic modes  in Fig.~\ref{fig:plane_cut_U=0}.
One can clearly see that the  cancellation of induced charges, which is required for the conservation of the total charge,
can happen locally (e.g., pairwise) or globally (e.g., in groups).
Some plasmons have simple patterns of dipolar, quadrupolar or triangular distribution of localized induced charges.
However, others have non-trivial symmetries and correspond to higher-order multipoles.

In Fig.~\ref{fig:plane_cut_U=2} we show the energy-resolved spatial distribution
of induced charges in impure graphene for a few selected plasmon energies
with characteristic patterns. 
These charge distributions are very different from those of the pristine lattice. We can clearly see that the  cancellation of 
induced charges can happen only over a larger area, because of the
 creation of  localized plasmons. Some of these modes have simple patterns of dipolar, quadrupolar, radial or triangular symmetry. However, others have highly non-trivial patterns, e.g., clumps of similar charges on nearest lattice sites.
The main result reported here is that the localization of plasmonic modes occurs at the nano-meter scale concentrating the 
electric field onto a very small area.


\section{Conclusions}

We investigated through exact diagonalization the plasmonic excitations of graphene in the presence of a single impurity with a special focus on localized modes (nano-plasmons) near the impurity. 
The dependence of excitations on the model parameters such as the sign, magnitude, size of the impurity potential and number of electrons in graphene was discussed. It was shown that the impurity potential and doping can be used to tune the properties of nano-plasmonic excitations, demonstrating that graphene is an inherently plasmonic material. 

Taking nano-plasmons in graphene as a case study, we demonstrated the importance of a three-dimensional, interactive visualization tool to explore and analyze large data sets in a more effective and time efficient way. 
We described the visualization problem faced with, even when the data set is only several Megabytes in size.
This issue arises when the data set is intrinsically multidimensional in both  real space and parameter space.
As a solution to this problem, we proposed a small-scale parallel processing approach, because it can no longer be efficiently handled by a single-processor computer.
Finally, using an interactive visualization tool we could extract even more quantitative information leading to
a better understanding of the physical properties of nano-plasmons on a two-dimensional lattice.

The results discussed can be tested, in principle, in experiments by high-frequency optical probes or local scanning tunneling spectroscopy probes. 
However, in order to achieve such tests, calculations of extended or nano-sized impurity clusters
are very desirable as they would give plasmonic excitations spread over a larger size with smaller frequencies.


\subsection*{Acknowledgements}
We are especially grateful to James Ahrens and John Patchett for many helpful discussions and support on the efficient use of ParaView.
This work was carried out under the auspices of the National Nuclear Security Administration of the U.S.\ Department of
Energy through the Laboratory Directed Research and Development program and the Office of Basic Energy Sciences at the
Center for Integrated Nanotechnologies at the Los Alamos National Laboratory 
under Contract No.\ DE-AC52-06NA25396 and Sandia National Laboratories under Contract No.\ DE-AC04-94AL85000.


 \end{document}